\documentclass[11pt,a4paper]{article}

\usepackage[english]{babel}
\usepackage[utf8x]{inputenc}
\usepackage[T1]{fontenc}

\usepackage[a4paper,top=3.5cm,bottom=3.5cm,left=3cm,right=3cm,marginparwidth=1.75cm]{geometry}

\usepackage{jheppub}
\usepackage{amsmath}
\usepackage{graphicx}
\usepackage{multirow}
\usepackage[colorinlistoftodos]{todonotes}

\newcommand{\ov}[1]{\overline{#1}}

\newcommand{\non}{\nonumber \\ }
\def\beq#1\eeq{\begin{align}#1\end{align}}

\numberwithin{equation}{section}

\title{{Probing SUSY effects in \boldmath ${K_S^0\rightarrow\mu^+\mu^-}$}}

\author[(a)]{Veronika Chobanova,}
\author[(b)]{Giancarlo D'Ambrosio,}
\author[(c,d)]{Teppei Kitahara,}
\author[(a)]{Miriam Lucio Mart\'inez,}
\author[(a)]{Diego Mart\'inez Santos,}
\author[(a)]{Isabel Su\'arez Fern\'andez}
\author[(e,f)]{\\ and Kei Yamamoto}

\affiliation[(a)]{Instituto Galego de F\'isica de Altas Enerx\'ias (IGFAE), Universidade de Santiago de Compostela, \\
R\'ua de Xoaqu\'in D\'iaz de R\'abago, s/n E-15782 Santiago de Compostela, Spain }
\affiliation[(b)]{INFN-Sezione di Napoli, \\
Via Cintia, 80126 Napoli, Italia}
\affiliation[(c)]{Institute for Theoretical Particle Physics (TTP), Karlsruhe Institute of Technology, \\ Engesserstra{\ss}e 7, D-76128 Karlsruhe, Germany}
\affiliation[(d)]{Institute for Nuclear Physics (IKP), Karlsruhe Institute of Technology, \\
Hermann-von-Helmholtz-Platz 1, D-76344 Eggenstein-Leopoldshafen, Germany}
\affiliation[(e)]{Department of Physics, Nagoya University, \\
Nagoya 464-8602, Japan}
\affiliation[(f)]{Kobayashi-Maskawa Institute for the Origin of Particles and the
Universe, Nagoya University, \\
Nagoya 464-8602, Japan}

\abstract{We explore supersymmetric contributions  to the decay $K_S^0\rightarrow\mu^+\mu^-$,
in light of
current experimental data. The Standard Model (SM) predicts $\mathcal{B}(K_S^0\rightarrow\mu^+\mu^-)\approx5\times 10^{-12}$. 
We find that contributions arising from flavour violating Higgs penguins can enhance the branching fraction up to  $\approx 35\times 10^{-12}$ within different scenarios of the Minimal Supersymmetric Standard Model (MSSM), as well as suppress it down to $\approx 0.78\times 10^{-12}$. Regions with fine-tuned parameters can bring the branching fraction up to the current experimental upper bound, $8\times 10^{-10}$. 
The mass degeneracy of the heavy Higgs bosons in MSSM induces correlations between $\mathcal{B}(K_S^0\rightarrow\mu^+\mu^-)$ and $\mathcal{B}(K_L^0\rightarrow\mu^+\mu^-)$. Predictions
for the $CP$ asymmetry in $K^0\rightarrow\mu^+\mu^-$ decays in the context of MSSM are also given, and can be up to eight times bigger than in the SM.}

\keywords{Rare kaon decays, Supersymmetry, direct $CP$ violation.}

\preprint{TTP17--048}

\begin{document}
\maketitle
\flushbottom


\thispagestyle{empty}
\newpage
\setcounter{page}{1}
\renewcommand{\thefootnote}{\#\arabic{footnote}}
\setcounter{footnote}{0}

\section{Introduction}

Leptonic decays of pseudoscalar mesons with down-type quarks are known to be very sensitive to the Higgs sector of the Minimal Supersymmetric Standard Model (MSSM), due to, among others, enhancement factors proportional to $\left(\tan^6\beta/M_A^4\right)$.\footnote{Note that this enhancement factor is not present in the up-type quark case.} 
This factor comes from the so-called non-holomorphic Yukawa terms at large $\tan \beta$ \cite{Hamzaoui:1998nu,Babu:1999hn,Chankowski:2000ng,Bobeth:2001sq,Isidori:2001fv,Isidori:2002qe},\footnote{
The higher-order contributions have been derived up to two-loop level in refs.~\cite{Crivellin:2010er, Crivellin:2011jt, Crivellin:2012zz}.}
which are triggered by the supersymmetric (SUSY) $\mu$ term, 
and hence the non-SUSY two-Higgs-doublet model cannot produce this enhancement \cite{Isidori:2001fv}.
The best known example is $B_s^0\rightarrow\mu^+\mu^-$ \cite{Hamzaoui:1998nu,Babu:1999hn,Chankowski:2000ng,Bobeth:2001sq,
Isidori:2001fv,Isidori:2002qe,Choudhury:1998ze,Huang:2000sm,Xiong:2001up,Dedes:2001fv,Bobeth:2002ch,Baek:2002rt,Dedes:2002zx, Mizukoshi:2002gs,Baek:2002wm}.
If Minimal Flavour Violation (MFV) is imposed,
then $B_s^0\rightarrow\mu^+\mu^-$ is the dominant constraint in $P\rightarrow\mu^+\mu^-$ decays. This is due to the stronger
Yukawa coupling of the $b$--quark compared to the $s$--quark, and to the better experimental precision in $B_s^0\rightarrow\mu^+\mu^-$ compared to $B_d^0\rightarrow\mu^+\mu^-$. 
However, in the presence of new sources of flavour violation, the sensitivity of each mode depends on the flavour and $CP$ structures of the corresponding terms.
Hence, a priori, $B_s^0\rightarrow\mu^+\mu^-$, $B_d^0\rightarrow\mu^+\mu^-$, $K_S^0\rightarrow\mu^+\mu^-$, and $K_L^0\rightarrow\mu^+\mu^-$ are all separate constraints that carry complementary information in the general MSSM. The observables related to these decay modes are typically branching fractions and $CP$ asymmetries. Even though the muon polarization could carry interesting information, it cannot be observed by current experiments.

In this paper, we focus on the MSSM effects in the $K_S^0\rightarrow\mu^+\mu^-$ decay. 
The Standard Model (SM) expectation is $(5.18\pm 1.50_{\textrm{LD}} \pm 0.02_{\textrm{SD}})\times 10^{-12}$~\cite{Ecker:1991ru, Isidori:2003ts, DAmbrosio:2017klp}, where the first uncertainty comes from the long-distance (LD) contribution and the second one comes from the short-distance (SD) contribution. 
On the other hand, the current experimental upper bound is $8\times10^{-10}$ at
$90\%$ C.L{.,} using $3$ fb${}^{-1}$ of LHCb data~\cite{LHCb:KsMuMu}. 
The LHCb upgrade could reach sensitivities at the level of about $1\times 10^{-11}$ or even below, approaching the SM prediction~\cite{DMS_FPCP}.

We predict the branching ratio $\mathcal{B}(K_S^0\rightarrow\mu^+\mu^-)$ under consideration of MSSM contributions and taking into account the relevant experimental constraints on the branching fractions
$\mathcal{B}(K_L^0\rightarrow\mu^+\mu^-)$, $\mathcal{B}(B^+\rightarrow\tau^+\nu_{\tau})$ and $\mathcal{B}(K^+\rightarrow\mu^+\nu_{\mu})$, the $CP$ violation parameters $\varepsilon^{\prime}_K/\varepsilon_K$ and $\varepsilon_K$, the $K^{0}_{L}$--$K^{0}_{S}$ mass difference, $\Delta M_{K}\equiv M_{K^0_L} - M_{K^0_S}  > 0$, and the Wilson coefficient $C_7$ from $b\rightarrow s \gamma$. 
We use the Mass Insertion Approximation (MIA)~\cite{MassInsertion}, treating the mass insertion terms as phenomenological parameters at the SUSY scale. The details of the formalism are given in section~\ref{sec:formalism}. The subsets of the MSSM parameter space are studied in scans performed on Graphics Processing Units (GPU), as detailed in section~\ref{sec:scan}. The results are shown in section~\ref{sec:results} and conclusions are drawn in section~\ref{sec:conclusions}.

\section{Formalism}
\label{sec:formalism}


\subsection{Definitions}
In this paper, 
we follow the notations of refs.~\cite{Altmannshofer:2009ne, Rosiek:1995kg}. 
We denote the right-handed down and up squarks as $D$ and $U$.
On the other hand,  the two left-handed squarks have the same mass because of the $\textrm{SU}(2)_{L}$ doublet, and they are denoted as $Q$.
The average of the $Q$, $D$, and $U$-squark masses squared are denoted by $\tilde{m}^2_Q$, $\tilde{m}^2_d$, $\tilde{m}^2_u$, respectively.

The mass insertions (hereafter MIs) are defined as: 
\beq
\left( \delta_d^{LL}\right)_{ij} &=
\frac{ \left[ \left( \mathcal{M}_D^2 \right)_{LL} \right]_{ij}  }{\tilde{m}_{Q}^2} 
=  \frac{ ( m^2_{Q})_{ji}}{\tilde{m}_{Q}^2},\\
\label{eq:duLL}
\left( \delta_u^{LL}\right)_{ij} &=
\frac{ \left[ \left( \mathcal{M}_U^2 \right)_{LL} \right]_{ij}  }{\tilde{m}_{Q}^2} 
= \frac{ ( V m^2_{Q} V^{\dag} )_{ji}}{\tilde{m}_{Q}^2},\\
\left( \delta_d^{RR}\right)_{ij} &=
\frac{ \left[ \left( \mathcal{M}_D^2 \right)_{RR} \right]_{ij}  }{\tilde{m}_{d}^2} 
=  \frac{ ( m^2_{D})_{ij}}{\tilde{m}_d^2},
\eeq
where $V$ is the Cabibbo–Kobayashi–Maskawa (CKM) matrix and $\mathcal{M}^{2}_{D,U}$ are the $6\times 6$ squark mass matrices.
Note that the indices $ij$ are inverted for $LL$.
Comparison with the SUSY Les Houches Accord 2 convention \cite{Allanach:2008qq} is given in the appendix of ref.~\cite{Altmannshofer:2009ne}.

The running coupling constants $\alpha_1$, $\alpha_2$, and $\alpha_3$ are defined as
\beq 
\label{eq:alphas}
\alpha_1 &= \frac{g_1^2}{4 \pi} = \frac{5}{3} \frac{g^{\prime 2}}{4 \pi}, \\
\alpha_2 &= \frac{g_2^2}{4 \pi} = \frac{g^2}{4 \pi}, \\
\alpha_3 &= \frac{g_3^2}{4 \pi} = \frac{g_s^2}{4 \pi},
\eeq 
where $g^{\prime}$, $g$, and $g_s$ are  the $\textrm{U}(1)_{Y}$, $\textrm{SU}(2)_{L}$, and  $\textrm{SU}(3)_{C}$ group coupling constants, respectively.	
In the following, these couplings are evaluated at the
$\mu^{\textrm{SUSY}}$ scale, where we define $\mu^{\textrm{SUSY}} = \sqrt{ \tilde{m}_{Q} M_3} $.

\subsection{Observables}
As will be shown in the next subsections, 
the main MSSM contribution to $\mathcal{B}(K_S^0\rightarrow\mu^+\mu^-)$ is proportional to $\left[\left(\delta_{d}^{LL(RR)}\right)_{12}\mu\tan^3\beta M_3/M_A^2\right]^2$. In order to constrain those parameters,
 the following observables are calculated in addition to $\mathcal{B}(K_S^0\rightarrow\mu^+\mu^-)$:
\begin{itemize}
\item Observables sensitive, among others, to the off-diagonal mass insertion terms $\left(\delta_{d}^{LL(RR)}\right)_{12}$: \\ $\mathcal{B}(K_L^0\rightarrow\mu^+\mu^-)$ , $\varepsilon^{\prime}_K/\varepsilon_K$, $\varepsilon_K$, and $\Delta M_{K}$.\footnote{
The contributions to $\mathcal{B}(K \to \pi \nu \overline{\nu})$ are controlled by an additional free parameter, the slepton mass, and $\mathcal{O}(1)$ effects are possible in this scenario  \cite{Crivellin:2017gks}.}
\item Observables sensitive to $\tan\beta$ and the heavy Higgs mass: 
$\mathcal{B}(B^+\rightarrow\tau^+\nu_{\tau})$, $\mathcal{B}(K^+\rightarrow\mu^+\nu_{\mu})$, $\Delta C_7$.
\end{itemize}

The definitions of $\mathcal{B}(B^+\rightarrow\tau^+\nu_{\tau})$, $\mathcal{B}(K^+\rightarrow\mu^+\nu_{\mu})$, and $C_{7}$ are given in ref.~\cite{Altmannshofer:2009ne} and the remaining observables are defined in the following subsections. The CKM matrix is fitted excluding measurements with potential sensitivity to MSSM contributions.

The constraints we impose on physics observables sensitive to the MSSM same parameters as $\mathcal{B}(K_S^0\rightarrow\mu^+\mu^-)$ are listed in table~\ref{tab:Observables}, 
where the EXP/SM represents the measured value over the SM prediction with their uncertainties. Due to the poor theoretical knowledge of $\Delta M_K$, we assign a $100\%$ theoretical uncertainty; thus, the constraint
imposed on this observable penalizes only $\mathcal{O}$(1) effects. It is not counted
as a degree of freedom in the $\chi^2$ tests, so that the $\Delta M_K$ constraint 
can only make the bounds tighter, but never looser.
Remaining constraints can in principle be satisfied by adjusting
the other parameters of the model. In particular, $B$ physics constraints not included in our list can be satisfied by parameters unspecified in our scan, for example by setting $\delta_{13} \approx \delta_{23} \approx 0$ and small $A_t$. The relation of eq.~\eqref{eq:duLL} may induce non-zero up-type MIs in the $B$ sector and hence modify $B^0_{s(d)}\rightarrow\mu^+\mu^-$, however, we
checked that these effects can be safely neglected in the scenarios we studied.
The large SUSY masses in our scan are typically beyond the reach of LHC.

The lattice values  for $(\varepsilon^{\prime}_K/\varepsilon_K)^{\textrm{SM}}$ used are from refs.~\cite{Blum:2011ng, Blum:2012uk,Blum:2015ywa,Bai:2015nea}, although
the conclusions of our study remain largely unchanged if we use the $\chi_{PT}$ value from 
refs.~\cite{Pallante:2001he,Hambye:2003cy,Mullor} instead. The values of  $\varepsilon_K^{\textrm{EXP}/\textrm{SM}}$ and $\Delta(\varepsilon^{\prime}_K/\varepsilon_K)^{{\rm EXP}-{\rm SM}}$ are discussed in more detail in the following subsections.

\begin{table}[!t]
\begin{center}
\begin{tabular}{c@{\hspace{0.05\textwidth}}c}
Observable & Constraint \\
\hline 
$\mathcal{B}(K_S^0\rightarrow\mu^+\mu^-)^{\rm EXP/SM}$ & unconstrained\\
\multirow{2}{*}{$\mathcal{B}(K_L^0\rightarrow\mu^+\mu^-)^{\rm EXP/SM}$} & $1.00 \pm 0.12$ (+)~\cite{DAmbrosio:2017klp, KLMuMu_theory,Patrignani:2016xqp}\\
 & $0.84 \pm 0.16$ ($-$)~\cite{DAmbrosio:2017klp,KLMuMu_theory,Patrignani:2016xqp} \\
\textbf{$\Delta M^{\rm EXP/SM}_{K}$} & $ 1\pm 1$ \\ 
\textbf{$\varepsilon_K^{\rm EXP/SM}$} & $ 1.05\pm 0.10$~\cite{Patrignani:2016xqp,Jang:2017ieg,Endo:2017ums} \\ 
\textbf{$\Delta(\varepsilon^{\prime}_K/\varepsilon_K)^{{\rm EXP}-{\rm SM}}$} & $ \left[ 15.5\pm 2.3 \text{(EXP)} \pm 5.07 \text{(TH)}\right] \times 10^{-4}$~\cite{Kitahara:2016nld, Patrignani:2016xqp} \\ 
\textbf{$\mathcal{B}(B^+ \rightarrow \tau^+ \nu_{\tau})^{\rm EXP/SM}$} & $0.91 \pm 0.22$~\cite{Patrignani:2016xqp} \\ 
\textbf{$\mathcal{B}(K^+ \rightarrow \mu^+ \nu_{\mu})^{\rm EXP/SM}$} & $1.0004 \pm 0.0095$~\cite{Patrignani:2016xqp} \\ 
\textbf{$\Delta C_7 $} & $-0.02 \pm 0.02$~\cite{C7_constraints} \\ 
$\tan\beta$:$M_{A}$ plane & ATLAS limits for hMSSM scenario~\cite{Aaboud:2017sjh} \\
LSP & Lightest neutralino \\ 
$B_G$ & $1 \pm 3\text{(TH)}$~\cite{Buras:1999da,Barbieri:1999ax} \\ 
\hline 
\end{tabular}
\end{center}
\caption{\label{tab:Observables}Physics observables constraints imposed in this study. The two different constraints on $\mathcal{B}(K_L^0\rightarrow\mu^+\mu^-)^{\rm EXP/SM}$  arise from an  unknown sign of  $A^{\mu}_{L\gamma\gamma}$ in eq.~\eqref{eq:ALgg}		~(see refs.~\cite{DAmbrosio:2017klp,KLMuMu_theory}).}
\end{table}

\subsection{$\boldsymbol{K^0 \to \mu^+ \mu^-}$}
\label{sec:KSmumu}
The $| \Delta S |= 1 $ effective Hamiltonian relevant for the $K^0 \rightarrow \ell \ov{\ell} $ transition at the $Z$ boson mass scale is
\beq
\mathcal{H}_{\rm eff} = - C_A Q_A - \tilde{C}_A \tilde{Q}_A  - C_S Q_S - \tilde{C}_S \tilde{Q}_S  - C_P Q_P - \tilde{C}_P \tilde{Q}_P + {\rm H.c.}, 
\eeq
where $C_{A}$, $C_{S}$ and $C_{P}$ are the axial, scalar and pseudoscalar Wilson coefficients. The right-handed and left-handed axial ($\tilde{Q}_A$, $Q_A$), scalar ($Q_S$, $\tilde{Q}_S$) and pseudoscalar ($Q_P$, $\tilde{Q}_P$) operators are given by:
\beq
Q_A & = (\overline{s} \gamma^{\mu} P_L d ) ( \overline{\ell} \gamma_{\mu} \gamma_5 \ell), ~~\tilde{Q}_A  = (\overline{s} \gamma^{\mu} P_R d ) ( \overline{\ell} \gamma_{\mu} \gamma_5 \ell),\non
Q_S &= m_s (\overline{s} P_R d) (\overline{\ell} \ell), ~~~~~~~\tilde{Q}_S = m_s (\overline{s} P_L d) (\overline{\ell} \ell),\non
Q_P &= m_s (\overline{s} P_R d) (\overline{\ell} \gamma_5 \ell), ~~~~\tilde{Q}_P = m_s (\overline{s} P_L d) (\overline{\ell} \gamma_5 \ell),
\eeq
where $P_{L,R}$ are the left and right-handed projection operators. 
For $\mathcal{B} ( K^0_{S,L} \rightarrow \mu^+ \mu^- )$~\footnote{
The electron modes are suppressed by $m_e^2 /m_{\mu}^2$, and we do not consider them in this paper.}, there are two contributions from S-wave ($A_{S,L}$) and P-wave transitions ($B_{S,L}$), resulting in:~\footnote{
Our result agrees with refs.~\cite{Mescia:2006jd,Altmannshofer:2011gn,Buras:2013uqa,Crivellin:2017upt}. However, it disagrees with notable literature \cite{Isidori:2002qe,Altmannshofer:2009ne} after discarding the long-distance contributions.
We found that  
$C_{10}^{{\rm SM}}$ should be $- C_{10}^{{\rm SM}}$ in eq.~(3.45) of ref.~\cite{Altmannshofer:2009ne}, and  $(C_P - C'_P)$ should be $(C'_P - C_P)$ in eq.~(2.4) of ref.~\cite{Isidori:2002qe}.
}
\beq
\mathcal{B}(K^0_{S,L} \to \mu^+ \mu^- ) =  \tau_{S,L} \Gamma (K^0_{S,L} \to \mu^+ \mu^- ) =  \tau_{S,L}  \frac{ f_K^2 M_K^3 \beta_{\mu}} { 16 \pi} 
  \left( |A_{S,L}|^2 + \beta_{\mu}^2 | B_{S,L} |^2\right),
  \label{eq:brKSLmm}
 \eeq
 with
\beq
A_S &=  \frac{ m_s M_K}{m_s + m_d} {\rm Im} ( C_P-\tilde{C}_P ) + \frac{2 m_{\mu}}{M_K} {\rm Im} (C_A  - \tilde{C}_A ),  \\
B_S &=  \frac{2 G_F^2 M_W^2 m_{\mu}  }{\pi^2 M_K}   B^{\mu}_{S \gamma \gamma} - \frac{ m_s M_K}{m_s + m_d}  {\rm  Re }(C_S - \tilde{C}_S),
\eeq
and
\beq
A_L &= \frac{2 G_F^2 M_W^2 m_{\mu}  }{\pi^2 M_K}  A^{\mu}_{L \gamma \gamma}    -  \frac{ m_s M_K}{m_s + m_d}  \textrm{Re}(C_P - \tilde{C}_P) -  \frac{2 m_{\mu}}{M_K}  \textrm{Re}(C_A  - \tilde{C}_A),\\ 
B_L &= \frac{ m_s M_K}{m_s + m_d}  \textrm{Im}(C_S - \tilde{C}_S),
 \label{eq:brKSLmmend}
\eeq
where
\beq
\beta_{\mu} =\sqrt{1 - \frac{4 m_{\mu}^2}{ M_K^2} }.
\eeq
Here, the long-distance contributions are \cite{Ecker:1991ru, Isidori:2003ts, DAmbrosio:2017klp, Mescia:2006jd}:  
\beq 
\frac{2 G_F^2 M_W^2 m_{\mu}  }{\pi^2 M_K}  B^{\mu}_{S \gamma \gamma} & =  (-2.65 + 1.14 i )\times 10^{-11} \textrm{\,(GeV})^{-2},\\
\frac{2 G_F^2 M_W^2 m_{\mu}  }{\pi^2 M_K}  A^{\mu}_{L \gamma \gamma} & = \pm   (0.54 - 3.96 i)\times 10^{-11} \textrm{\,(GeV})^{-2},
\label{eq:ALgg}
\eeq 
with\footnote{
Note that 
$B^{\mu}_{S \gamma \gamma}$ is denoted by $A^{\mu}_{S \gamma \gamma}$ in refs.~\cite{DAmbrosio:2017klp, Mescia:2006jd}.}
\beq
B^{\mu}_{S \gamma \gamma} &= \frac{\pi \alpha_0}{ G_F^2 M_W^2 f_K {M_{K}} |H(0)|} \mathcal{I} \left( \frac{m_{\mu}^2}{M_K^2}, \frac{m_{\pi^{\pm}}^2}{M_K^2} \right)  \sqrt{ \frac{2 \pi}{{M_{K}} } \frac{\mathcal{B}(K_S^0 \to \gamma \gamma)^{\rm EXP}}{\tau_S} }, \\
 {A}^{\mu}_{L\gamma \gamma} 
  &= \frac{ \pm  2 \pi \alpha_0}{G_F^2 M_W^2 f_K M_{K}} \mathcal{A}\left(M_K^2\right)\sqrt{ \frac{  2 \pi }{{M_{K}}}  \frac{\mathcal{B}(K_L^0 \to \gamma \gamma)^{\rm EXP}}{\tau_L}}  
    ,
\eeq
where
a two-loop function $\mathcal{I} (a,b)$ from the $2\pi^{\pm} 2 \gamma$  intermediate state  is given in refs.~\cite{Isidori:2004rb, Ecker:1991ru},
a pion one-loop contribution with  two external on-shell photons is represented as $H(0) = 0.331+ 
 i 0.583$ \cite{Ecker:1991ru}, and  
a one-loop function  $\mathcal{A} (s)$ from the $ 2 \gamma$ intermediate state  is 
  given in refs.~\cite{GomezDumm:1998gw,Knecht:1999gb}.
Here, $\alpha_0 = 1/137.04$,  $f_K = ( 155.9 \pm 0.4 )$ MeV \cite{Patrignani:2016xqp}, and $\tau_{S,L}$ are the $K^0_{S,L}$ lifetimes. 
Note that there is a theoretically and experimentally unknown sign in $A^{\mu}_{L \gamma \gamma}$, 
which is determined by higher chiral orders than $\mathcal{O}(p^4)$ contributions \cite{Pich:1995qp,Gerard:2005yk}, and they provide two different constraints on $\mathcal{B}(K_L^0\rightarrow\mu^+\mu^-)^{\rm EXP/SM}$ in table~\ref{tab:Observables}.
This sign can be determined by a precise measurement of the interference between $K^0_{L} \to \mu^+ \mu^-$ and $K^0_{S} \to \mu^+ \mu^-$ \cite{DAmbrosio:2017klp}.
In addition, in the MSSM, 
the correlation between $\mathcal{B}(K^0_{S} \to \mu^+ \mu^- )$
and $\mathcal{B}(K^0_{L} \to \mu^+ \mu^- )$ depends on the unknown sign of $A^{\mu}_{L \gamma \gamma} $.
In the following, we derive some relations between the two branching fractions, for a better interpretation of the results of our scans.
In the case in which new physics enters only in
$\tilde{C}_S$ and $\tilde{C}_P = \tilde{C}_S$ (pure left-handed MSSM scenario),  
the following relations between the branching fractions of $K^0_S$ and $K_L^0$ decaying into $\mu^+ \mu^-$ can be established:
\beq 
\label{eq:FF1}
\mathcal{B}\left(K^0_S \to \mu^+ \mu^- \right) \propto & \beta_{\mu}^2 \left| N^{\rm LD}_{S} \right|^2 + \left(A^{{\rm SD}}_{S, {\rm SM}}\right)^2  
- 2 M_K \left[  A^{{\rm SD}}_{S, {\rm SM}} \textrm{Im}  (\tilde{C}_S) -  \beta_{\mu}^2  \textrm{Re} \left( N^{\rm LD}_{S} \right)  \textrm{Re}  (\tilde{C}_S)  \right]\non
 & + M_K^2 \left\{ \left[ \textrm{Im}  (\tilde{C}_S) \right]^2 + \beta_{\mu}^2 \left[ \textrm{Re}  (\tilde{C}_S) \right]^2 \right\},\\
\mathcal{B}\left(K^0_L \to \mu^+ \mu^- \right)  \propto &  \left| N^{\rm LD}_{L} \right|^2 +  \left(A^{{\rm SD}}_{L, {\rm SM}}\right)^2    
- 2 M_K  \textrm{Re}  (\tilde{C}_S) \left[ A^{{\rm SD}}_{L, {\rm SM}}  - \textrm{Re} \left( N^{\rm LD}_{L} \right)\right] \non
& + M_K^2 \left\{ \left[ \textrm{Re}  (\tilde{C}_S) \right]^2 + \beta_{\mu}^2 \left[ \textrm{Im}  (\tilde{C}_S) \right]^2 \right\} - 2  A^{{\rm SD}}_{L, {\rm SM}} \textrm{Re} \left( N^{\rm LD}_{L} \right),
\eeq
with
\beq
A^{{\rm SD}}_{S, {\rm SM}} = \frac{2 m_{\mu}}{M_K} \textrm{Im}(C_{A, {\rm SM}}), \quad
A^{{\rm SD}}_{L, {\rm SM}} = \frac{2 m_{\mu}}{M_K} \textrm{Re}(C_{A, {\rm SM}}),
\eeq
and 
\beq
N^{\rm LD}_S =  \frac{2 G_F^2 M_W^2 m_{\mu}  }{\pi^2 M_K}  B^{\mu}_{S \gamma \gamma}, \quad
N^{\rm LD}_L =  \frac{2 G_F^2 M_W^2 m_{\mu}  }{\pi^2 M_K}  A^{\mu}_{L \gamma \gamma},
\eeq
where $m_d$ terms are discarded for simplicity.
The long-distance term $\textrm{Re} \left( N^{\rm LD}_L \right)$ holds the unknown sign from $  A^{\mu}_{L \gamma \gamma}$, which changes the correlation significantly, as will be shown. 
On the other hand, if new physics produces only
$C_S$ and $C_P = -C_S$ (pure right-handed MSSM),
the two branching fractions are

\beq 
\mathcal{B}\left(K^0_S \to \mu^+ \mu^- \right) \propto & \beta_{\mu}^2 \left| N^{\rm LD}_{S} \right|^2 + \left(A^{{\rm SD}}_{S, {\rm SM}}\right)^2  - 2 M_K \left[  A^{{\rm SD}}_{S, {\rm SM}} \textrm{Im}  (C_S) + \beta_{\mu}^2  \textrm{Re} \left( N^{\rm LD}_{S} \right)  \textrm{Re}  (C_S)  \right]\non
 & + M_K^2 \left\{ \left[ \textrm{Im}  (C_S) \right]^2 + \beta_{\mu}^2 \left[ \textrm{Re}  (C_S) \right]^2 \right\},\\
\mathcal{B}\left(K^0_L \to \mu^+ \mu^- \right)  \propto &  \left| N^{\rm LD}_{L} \right|^2 +  \left(A^{{\rm SD}}_{L, {\rm SM}}\right)^2    
- 2 M_K  \textrm{Re}  (C_S) \left[ A^{{\rm SD}}_{L, {\rm SM}}  - \textrm{Re} \left( N^{\rm LD}_{L} \right)\right] \non
& + M_K^2 \left\{ \left[ \textrm{Re}  (C_S) \right]^2 + \beta_{\mu}^2 \left[ \textrm{Im}  (C_S) \right]^2 \right\} - 2  A^{{\rm SD}}_{L, {\rm SM}} \textrm{Re} \left( N^{\rm LD}_{L} \right).
\label{eq:FF2}
\eeq
It is shown that $ \mathcal{B}\left(K^0_L \to \mu^+ \mu^- \right)$ is the same as the pure left-handed one by a replacement of $C_S \to \tilde{C}_S$, while $ \mathcal{B}\left(K^0_S \to \mu^+ \mu^- \right)$ is not; the final terms of the first line have opposite sign. Hence, the relations
between the two branching fractions are different for left-handed and
right-handed new physics scenarios.

 
For those cases, the experimental measurement of $\mathcal{B}(K_L^0\rightarrow\mu^+\mu^-)$ \cite{Patrignani:2016xqp}, 
\beq
\mathcal{B}(K_L^0\rightarrow\mu^+\mu^-)^{\text{EXP}} = \left(6.84 \pm  0.11 \right) \times 10^{-9},
\label{eq:KLmmexp}
\eeq
imposes an upper bound on $\mathcal{B}(K_S^0\rightarrow\mu^+\mu^-)$. 
This bound can be alleviated if $|C_S|\neq|C_P|$ or if new physics is present
simultaneously in the left-handed and right-handed Wilson coefficients.

Experimentally, one can also access an {\it effective} branching ratio of $K_S^0 \to  \mu^+ \mu^-$~\cite{DAmbrosio:2017klp} which includes an interference contribution with $K_L^0 \to  \mu^+ \mu^-$ in the neutral kaon sample.
We obtain  
\beq
&\mathcal{B} (K_S^0 \to \mu^+ \mu^-)_{\rm eff}=\tau_S  \left( \int^{t_{{\rm max}}}_{t_{{\rm min}}} d t e^{- \Gamma_S t} \varepsilon (t)\right)^{-1}
 \Biggl[ \int^{t_{{\rm max}}}_{t_{{\rm min}}} d t \Biggl\{  \Gamma (K^0_{S} \to \mu^+ \mu^- )  e^{- \Gamma_S t}   \non
& ~~+   \frac{ D f_K^2 M_K^3 \beta_{\mu}}{ {8} \pi} \textrm{Re}\left[  i \left( A_S A_L - \beta_{\mu}^2 {B_S^{\ast}} B_L \right) e^{ - i \Delta M_K t}  \right] e^{- \frac{ \Gamma_S + \Gamma_L}{2} t } \Biggr\} \varepsilon (t) \Biggr], 
\label{eq:effBR}
\eeq
where the dilution factor $D$ is a measure of the initial ($t=0$) $K^0$--$\overline{K}{}^0$  asymmetry,
\begin{align}
 D= \frac{ K^0 - \overline{K}{}^0 }  { K^0 + \overline{K}{}^0 },
 \label{eq:DE}
\end{align}
 $\varepsilon(t)$ is the decay-time acceptance of the detector. 
The second line of eq.~\eqref{eq:effBR} corresponds to an interference effect between $K_L^0$ and $K_S^0$, and for $D= 0 $, $\mathcal{B} (K_S^0 \to \mu^+ \mu^-)_{\rm eff}$ corresponds to  $\mathcal{B} (K_S^0 \to \mu^+ \mu^-)$. 
The current experimental bound 
\cite{LHCb:KsMuMu}, 
\beq
\mathcal{B}(K_S^0\rightarrow\mu^+\mu^-)^{\text{EXP}} < 8  \times 10^{-10} ~[90\%~{\rm C.L.}],
\eeq
uses untagged $K^0$ and $\bar{K^0}$ mesons produced in almost equal amounts, and hence $D=0$ is assumed. A pure $K_L^0 \rightarrow \mu^+\mu^-$ background can be subtracted by a combination of  simultaneous measurement of $K^0_S \to \pi^+ \pi^- $ events and knowledge of the observed value of $\mathcal{B}(K_L^0\rightarrow\mu^+\mu^-)$ in eq.~\eqref{eq:KLmmexp} \cite{DAmbrosio:2017klp}. 
The decay-time acceptance of the LHCb detector is parametrized by
$\varepsilon(t) = \exp( - \beta t)$ with $\beta \simeq 86\,$ns$^{ -1}$,
and the range of the detector for selecting $K^0 \to \mu^+ \mu^-$ is $t_{{\rm min}} = 8.95\,$ps$ = 0.1 \tau_S$ and $t_{{\rm max}} = 130\,$ps $= 1.45 \tau_S$.

Given the potential measurement of an effective branching ratio by different dilution factors $D >0 $ and $D' < 0$ using $K^-$ tagging and $K^+$ tagging \cite{DAmbrosio:2017klp}, respectively, the direct $CP$ asymmetry can be measured using the difference 
$
\mathcal{B} (K_S^0 \to \mu^+ \mu^-)_{\rm eff}(D)  - \mathcal{B} (K_S^0 \to \mu^+ \mu^-)_{\rm eff}(D')
$,
which is a theoretically clean quantity that emerges from a genuine direct $CP$ violation.
Here, the charged kaon is accompanied by the neutral kaon beam as, for instance, $pp \to K^0 K^- X$ or $pp \to \overline{K}{}^0 K^+ X$.
Note that a definition of $D'$ is the same as $D$ in eq.~\eqref{eq:DE} but charged kaons of opposite sign are  required in the event selection. 
 Therefore, we define the following direct $CP$ asymmetry in $K^0_S \to \mu^+ \mu^-$:
 \beq
 A_{CP} (K^0_S \to \mu^+ \mu^-)_{D,D'} = 
\frac{\mathcal{B} (K^0_S \to \mu^+ \mu^-)_{\rm eff}(D)  - \mathcal{B} (K^0_S \to \mu^+ \mu^-)_{\rm eff}(D')}
{\mathcal{B} (K^0_S \to \mu^+ \mu^-)_{\rm eff}(D)  + \mathcal{B} (K^0_S \to \mu^+ \mu^-)_{\rm eff}(D')}.
\eeq
We discarded the indirect $CP$-violating contributions because they are numerically negligible compared to the $CP$-conserving and the direct $CP$-violating  contributions \cite{DAmbrosio:2017klp}.

Within the SM, the Wilson coefficients are, 
\beq
C_{A, {\rm SM}} &= - \frac{\left[ \alpha_2(M_Z)\right]^2}{2 M_W^2}    \left(V_{ts}^{\ast} V_{td}  Y_t + V_{cs}^{\ast} V_{cd} Y_{c} \right), \\
\tilde{C}_{A, {\rm SM}} &= C_{S, {\rm SM}} = \tilde{C}_{S, {\rm SM}} = C_{P, {\rm SM}} = \tilde{C}_{P, {\rm SM}} \simeq 0,
\eeq
where $Y_t = 0.950  \pm  0.049 $ and $Y_c = ( 2.95 \pm 0.46 ) \times 10^{-4}$ \cite{Gorbahn:2006bm}.
Using the CKM matrix tailored for probing the MSSM contributions, we obtain the SM prediction of $A_{CP}$,
\beq
A_{CP} (K^0_S \to \mu^+ \mu^-)_{D,D'}^{\rm SM}
= \left\{ 
\begin{array}{ll}
- \frac{3.71 \left( D - D' \right) }{ \left(10.53 \pm 3.01 \right) - 3.71 \left( D+D' \right) }, & (+) \\
\frac{3.98 \left( D - D' \right) }{ \left(10.53 \pm 3.01 \right) + 3.98 \left( D+D' \right) }, & (-)
\end{array}
\right.
\eeq
where $(+)$ and $(-)$ correspond to the unknown sign of $A^{\mu}_{L \gamma \gamma}$ in eq.~\eqref{eq:ALgg}. The uncertainty is  totally dominated by $B^{\mu}_{S \gamma \gamma}$ \cite{DAmbrosio:2017klp} and it will be sharpened by the dispersive treatment of $K_S^0 \to \gamma^{({\ast})} \gamma^{({\ast})} $ \cite{Colangelo:2016ruc}.
If one considers the case of $D' = -D$ achieved by the accompanying opposite-charged-kaon tagging, the SM prediction of $A_{CP}$ is simplified: 
\beq
A_{CP} (K^0_S \to \mu^+ \mu^-)_{D, -D}^{\rm SM}
=\left\{ 
\begin{array}{ll}   \left( - 0.704 {}_{- 0.281}^{+0.156} \right) \times D, & (+) \\
\left(+0.756 {}^{+0.302}_{-0.168} \right) \times D. & (-) 
\end{array}
\right.
\label{sq:ACPSM}
\eeq

In the MSSM, 
the leading contribution to $C_A$, induced by terms of second order in the expansion of the squark mass matrix of the chargino $Z$-penguin, is \cite{ Isidori:2002qe, Colangelo:1998pm}, 
\beq
C_A&=- \frac{ (\alpha_2)^2}{16 M_W^2}
\frac{ \left[ (\mathcal{M}_U^2)_{LR} \right]^{\ast}_{2 3 } \left[ (\mathcal{M}_U^2)_{LR}\right]_{ 1 3} }{M_2^4} l \left( x^Q_2 , x^u_2 \right),\\
\tilde{C}_A &=0,
\eeq
where $x^Q_2 =  \tilde{m}^2_Q /M_2^2 $ and $ x^u_2 = \tilde{m}^2_{u} / M_2^2 $. 
The loop function $l(x,y)$ \cite{Colangelo:1998pm} is defined in  
appendix~\ref{App:loopKmm}.
Here, contributions from the Wino-Higgsino mixing  are omitted.
Setting $\tilde{m}^2_Q  = \tilde{m}^2_{u}$ gives the MIA result of refs.~\cite{Buras:1999da,Endo:2016aws}.

\begin{figure}
\centering
\includegraphics[width=0.5\textwidth]{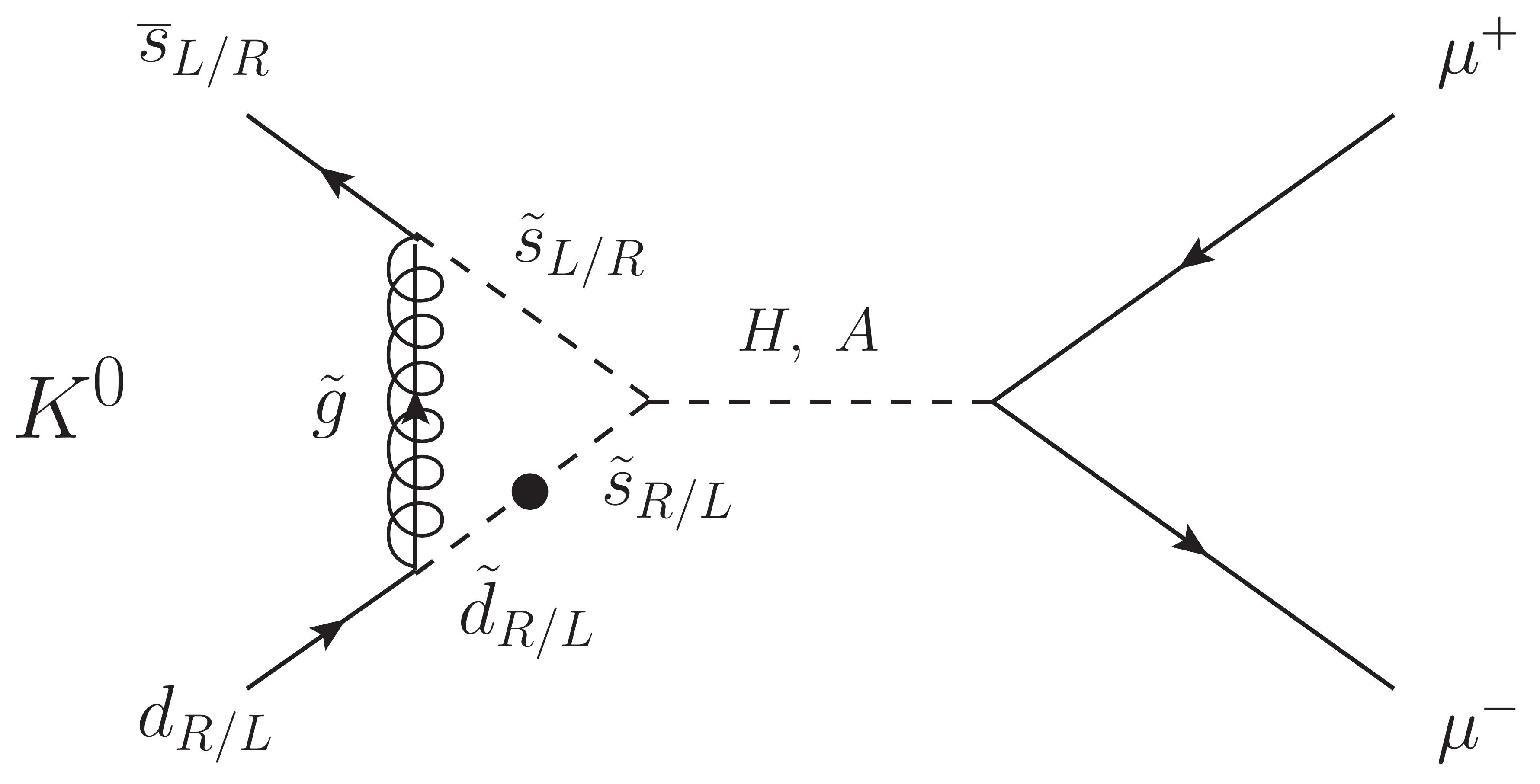}
\caption{Feynman diagram of the leading (pseudo-)scalar MSSM contributions to $K^0_S\rightarrow\mu^+\mu^-$ and $K^0_L \rightarrow\mu^+\mu^-$, which include a gluino and a heavy Higgs boson. 
The black dot is the corresponding mass insertion term.}
\label{fig:feyn_gluino}
\end{figure}

The leading MSSM contributions to $C_{S(P)}$ and $\tilde{C}_{S(P)}$ in $K^0_S\rightarrow\mu^+\mu^-$ and $K^0_L \rightarrow\mu^+\mu^-$ are shown in figure~\ref{fig:feyn_gluino}. 
For $C_S$ and $\tilde{C}_S$,   we obtain
\beq
{C}_S =& 
- \frac{ 2}{3} \frac{ \alpha_s \alpha_2  m_{\mu}}{ M_W^2} \frac{ \mu M_3}{M_A^2 \tilde{m}_{d}^2} \left( \delta^{RR}_{d} \right)_{12} \frac{\tan^3 \beta}{ (1 + \epsilon_g \tan \beta)^2 (1 + \epsilon_{\ell} \tan \beta)} G \left( x^3_d, x^Q_d  \right) \non
& - \frac{ 2}{3} \frac{ \alpha_s \alpha_2  m_{\mu}}{ M_W^2} \frac{m_b}{m_s}  \frac{ \mu M_3 \tilde{m}_{Q}^2}{M_A^2 \tilde{m}_{d}^4} \left( \delta^{RR}_{d} \right)_{13}  \left( \delta^{LL}_{d} \right)_{32} \non
& ~~~~ \times  \frac{\tan^3 \beta}{ (1 + \epsilon_g \tan \beta)[1 + (\epsilon_g  + \epsilon_Y y_t^2 )\tan \beta] (1 + \epsilon_{\ell} \tan \beta)} H \left( x^3_d, x^Q_d  \right),\\
\tilde{C}_S =& 
- \frac{ 2}{3} \frac{ \alpha_s \alpha_2  m_{\mu}}{ M_W^2} \frac{ \mu M_3}{M_A^2 \tilde{m}_{Q}^2} \left( \delta^{LL}_{d} \right)_{12} \frac{\tan^3 \beta}{ (1 + \epsilon_g \tan \beta)^2 (1 + \epsilon_{\ell} \tan \beta)} G \left( x^3_Q, x^d_Q \right) \non
& - \frac{ 2}{3} \frac{ \alpha_s \alpha_2  m_{\mu}}{ M_W^2} \frac{m_b}{m_s}  \frac{ \mu M_3 \tilde{m}_{d}^2}{M_A^2 \tilde{m}_{Q}^4} \left( \delta^{LL}_{d} \right)_{13}  \left( \delta^{RR}_{d} \right)_{32} \non
& ~~~~ \times  \frac{\tan^3 \beta}{ (1 + \epsilon_g \tan \beta)[1 + (\epsilon_g  + \epsilon_Y y_t^2 )\tan \beta] (1 + \epsilon_{\ell} \tan \beta)} H \left( x^3_Q, x^d_Q \right)\non
& + \frac{(\alpha_2)^2 m_{\mu} m_t^2 }{8 M_W^4}
\frac{ \mu A_t}{ M_A^2 \tilde{m}_{Q}^2} V_{ts}^{\ast} V_{td}  \frac{\tan^3 \beta [ 1 + ( \epsilon_g + \epsilon_Y y_t^2 )\tan \beta]^2 }{ (1 + \epsilon_g \tan \beta )^4 (1  + \epsilon_{\ell} \tan \beta ) }  F \left( x^{\mu}_Q,x^{u}_Q \right)\non
&+  \frac{(\alpha_2)^2 m_{\mu} }{ 4 M_W^2} \frac{ \mu M_2}{M_A^2\tilde{m}_{Q}^2}  \left( \delta^{LL}_u \right)_{12} \frac{ \tan^3 \beta}{ (1 + \epsilon_g  \tan \beta  )^2 (1  + \epsilon_{\ell} \tan \beta )} 
G\left( x^2_Q, x^{\mu}_Q \right),
\eeq
with 
\beq
\epsilon_g &= \frac{ 2 \alpha_s}{ 3 \pi} \frac{ \mu M_3}{ \tilde{m}_{Q}^2 } F\left( x^3_Q, x^d_Q \right),\\
\epsilon_Y & = \frac{1}{16 \pi} \frac{ \mu A_t}{ \tilde{m}_{Q}^2} F\left( 
x^{\mu}_Q, x^{u}_Q \right),\\
\epsilon_{\ell} &\simeq - \frac{ 3 \alpha_2}{ 16 \pi},
\eeq
where $x^3_d =   M_3^2/ \tilde{m}_{d}^2$, $x^Q_d = \tilde{m}_{Q}^2/ \tilde{m}_{d}^2 $, $x^3_Q =   M_3^2/ \tilde{m}_{Q}^2$, 
$x^d_Q =\tilde{m}_{d}^2 / \tilde{m}_{Q}^2$, 
$x^{\mu}_Q = \mu^2 / \tilde{m}_{Q}^2$,
$x^{u}_Q = \tilde{m}_{u}^2 / \tilde{m}_{Q}^2$,
$x^{2}_Q = M_2^2 / \tilde{m}_{Q}^2$, and 
$x^{\mu}_Q = \mu^2 / \tilde{m}_{Q}^2$.
The loop functions $F(x,y)$, $G(x,y)$, and $H(x,y)$ are defined in appendix~\ref{App:loopKmm}.
These results are consistent with ref.~\cite{Altmannshofer:2009ne} in the universal squark mass limit after changing the flavour and its chirality for $B_s^0$ decay.
Here, we used the following approximation 
\beq
 \alpha \simeq \beta-\frac{\pi}{2}, \quad
 M_{H}\simeq  M_{A},
 \label{eq:appMA}
\eeq
where $\alpha$ is an angle  of the orthogonal rotation matrix for the $CP$-even Higgs mass, and $M_{H}$ ($M_{A}$) is a $CP$-even (odd) heavy Higgs mass.
On the other hand, the contributions to $C_P$ and $\tilde{C}_P$ are
\beq
C_P = - C_S, \quad
\tilde{C}_P = \tilde{C}_S.
\eeq
Note that the Wilson coefficients in the MSSM are given at the $\mu^{\rm SUSY} $ scale, and there is no QCD correction from the renormalization-group (RG) evolution at the leading order.

\subsection{$\boldsymbol{\varepsilon^{\prime}_K / \varepsilon_K}$}

New physics models affecting $\varepsilon^{\prime}_K/\varepsilon_K$ have
recently attracted some attention
since  lattice  results from the RBC and UKQCD collaborations~\cite{Blum:2011ng, Blum:2012uk,Blum:2015ywa,Bai:2015nea} have been reported 
$2$--$3\sigma$ below \cite{Buras:2015yba,Kitahara:2016nld} the experimental world average of Re$(\varepsilon^\prime_K / \varepsilon_K)$ \cite{Patrignani:2016xqp}.
This is consistent  with  the recent calculations in the large-$N_c$ analyses~\cite{Buras:2015xba,Buras:2016fys}. 
Although the lattice simulation \cite{Bai:2015nea}  includes final-state interactions partially along the line of ref.~\cite{Lellouch:2000pv},
final-state interactions have to be still fully included in the calculations in light of a discrepancy of a strong phase shift $\delta_0$~\cite{Colangelo:2001df,GarciaMartin:2011cn,Colangelo:NA62}.  
Conversely combining large-$N_c$ methods with chiral loop corrections can bring the value of $\varepsilon^{\prime}_K/\varepsilon_K$ in agreement with the experiment~\cite{Pallante:2001he,Hambye:2003cy,Mullor}.

In this paper, we used the hadronic matrix elements obtained by lattice simulations.
For the $\chi^2$ test, we use the following constraint,
\beq
\Delta \left( \frac{\varepsilon^{\prime}_K}{\varepsilon_K}\right)^{{\rm EXP}-{\rm SM}} \equiv 
\textrm{Re}\left( \frac{\varepsilon^{\prime}_K}{\varepsilon_K}\right)^{\rm EXP} - \left( \frac{\varepsilon^{\prime}_K}{\varepsilon_K}\right)^{\rm SM} = 
\left[ 15.5\pm 2.3 \text{(EXP)} \pm 5.07 \text{(TH)}\right] \times 10^{-4},
\eeq
with
\beq
\left( \frac{\varepsilon^{\prime}_K}{\varepsilon_K}\right)^{\rm SM} \rightarrow \left( \frac{\varepsilon^{\prime}_K}{\varepsilon_K}\right)^{\rm SM} + \left( \frac{\varepsilon^{\prime}_K}{\varepsilon_K}\right)^{\rm SUSY},
\eeq
where the SM prediction at the next-to-leading order in ref.~\cite{Kitahara:2016nld} is used. The experimental value of $\varepsilon_K$ is used in the calculation of the ratio. The SUSY contributions to $\varepsilon_K$ are given in the next subsection.

Within the MSSM, the SUSY contributions to  $\varepsilon^{\prime}_K/ \varepsilon_K $ are dominated by
gluino box, chargino-mediated $Z$-penguin, and chromomagnetic dipole contributions.
The first two contributions are represented by the same $| \Delta S |= 1 $ four-quark effective Hamiltonian at the $\mu^{\textrm{SUSY}}$ scale, which is: 
\beq
\mathcal{H}_{\rm eff} = \frac{G_F}{\sqrt{2} }\sum_{q} \sum_{i=1}^{4} \left[ C_i^q Q_{i}^q + \tilde{C}_{i}^{q} \tilde{Q}_{i}^{q} \right] + {\rm H.c.}, 
\eeq
with
\beq
&Q^{q}_1 = \left( \bar{s}  d \right)_{V-A} \left( \bar{q}  q \right)_{V+A},
~~~~~~~~\tilde{Q}^{q}_1 = \left( \bar{s}  d \right)_{V+A} \left( \bar{q}  q \right)_{V-A},\non
&Q^{q}_2 = \left( \bar{s}_{\alpha}  d_{\beta} \right)_{V-A} \left( \bar{q}_{\beta}  q_{\alpha} \right)_{V+A},~~
\tilde{Q}^{q}_2 = \left( \bar{s}_{\alpha}  d_{\beta} \right)_{V+A} \left( \bar{q}_{\beta}  q_{\alpha} \right)_{V-A},  \non
&Q^{q}_3 = \left( \bar{s}  d \right)_{V-A} \left( \bar{q}  q \right)_{V-A},
~~~~~~~~\tilde{Q}^{q}_3 = \left( \bar{s}  d \right)_{V+A} \left( \bar{q}  q \right)_{V+A},\non
&Q^{q}_4 = \left( \bar{s}_{\alpha}  d_{\beta} \right)_{V-A} \left( \bar{q}_{\beta}  q_{\alpha} \right)_{V-A},~~
\tilde{Q}^{q}_4 = \left( \bar{s}_{\alpha}  d_{\beta} \right)_{V+A} \left( \bar{q}_{\beta}  q_{\alpha} \right)_{V+A}, 
\eeq
where $(V \mp A)$ refers to $\gamma_{\mu} (1 \mp \gamma_5)$, and $\alpha$ and $\beta$
 are color indices.
 
The Wilson coefficients from the gluino box contributions are leading contributions when the mass difference between right-handed squarks exists \cite{Kitahara:2016otd, Kagan:1999iq}. They are shown in appendix~\ref{App:WCgluino} with their corresponding loop functions defined in appendix~\ref{App:gluinobox}. Here,  $(\delta_d)_{13} (\delta_d)_{32}$ terms are discarded for simplicity.

The Wilson coefficients of the chargino-mediated $Z$-penguin are induced by terms of second order in the expansion of MIA. These ones are shown in appendix~\ref{App:WCchargino}, where the loop function  $l(x,y)$ is given by eq.~\eqref{LoopL}.

The matching conditions to the standard four-quark Wilson coefficients \cite{Kitahara:2016nld} are

\beq 
\begin{array}{ll}
s_1 = 0, & s_2 = 0, \\
\\
s_3    =  \frac{1}{3} \left( C_3^{u}  + 2 C_3^{d}   \right), & s_4    = \frac{1}{3}  \left( C_4^{u}  + 2 C_4^{d}   \right), \\
\\
s_5    =  \frac{1}{3} \left( C_1^{u}  + 2 C_1^{d}   \right), & s_6    = \frac{1}{3} \left( C_2^{u}  + 2 C_2^{d}   \right), \\
\\
s_7    =  \frac{2}{3}\left( C_1^{u}  - C_1^{d}   \right), & s_8    = \frac{2}{3}  \left( C_2^{u}  - C_2^{d}   \right), \\
\\
s_9    =  \frac{2}{3} \left( C_3^{u}  - C_3^{d}   \right), & s_{10}    =  \frac{2}{3}  \left( C_4^{u}  - C_4^{d}   \right).
\end{array}
\eeq
The coefficients for the opposite-chirality operators, $\tilde{s}_{1,\dots, 10}$, are trivially found from the previous ones by
replacing ${C}_{1,2,3,4}^{q} \rightarrow \tilde{C}_{1,2,3,4}^{q} $.
Using the Wilson coefficients $\vec{s} = (s_1, s_2, \dots, s_{10})^{T}$ and 
$\vec{\tilde{s}} = (\tilde{s}_1, \tilde{s}_2, \dots, \tilde{s}_{10})^{T}$ at the $\mu^{\textrm{SUSY}}$ scale, 
the dominant box and penguin contributions to $\varepsilon^{\prime}_K / \varepsilon_K$ are  given by \cite{Kitahara:2016nld}
\beq
\left. \frac{\varepsilon^{\prime}_K}{\varepsilon_K} \right|_{\textrm{box}+\textrm{pen}}
& = \frac{G_F \omega_{+}}{ 2 |\varepsilon_K^{\textrm{EXP}}| \textrm{Re}A_0^{\textrm{EXP}}}  \langle \vec{Q}_{\varepsilon^{\prime}} (\mu )^{T} \rangle
\hat{U}(\mu, \mu^{\textrm{SUSY}}) \textrm{Im} \left[ \vec{s} - 
\vec{\tilde{s}}    \right],
\eeq
with
\beq
\omega_{+ } & = \left( 4.53 \pm 0.02 \right) \times 10^{-2},\\
|\varepsilon_K^{\textrm{EXP}}| &=  \left( 2.228 \pm 0.011 \right) \times 10^{-3}, \label{eq:epKEXP}\\
\textrm{Re}A_0^{\textrm{EXP}} & =  \left( 3.3201 \pm 0.0018 \right) \times  10^{-7} \textrm{~GeV}.
\eeq
The hadronic matrix elements at $\mu = 1.3$ GeV, including $I=0$ and $I=2$ parts, are \cite{Kitahara:2016nld}
\beq
\langle \vec{Q}_{\varepsilon^{\prime}} (\mu  )^{T} \rangle =& \Bigl(  0.345, 0.133, 0.034, -0.179, 0.152, 0.288, 2.653, 17.305, 0.526, 0.281 \Bigr) ~\textrm{(GeV})^3, 
\eeq
and the approximate function of the RG evolution matrix $\hat{U}(\mu, \mu^{\textrm{SUSY}})$ is given in ref.~\cite{Kitahara:2016nld}.

Next, the $| \Delta S |= 1 $ chromomagnetic-dipole operator that contributes to $\varepsilon^{\prime}_K / \varepsilon_K$ 
is 
\beq
\mathcal{H}_{\textrm{eff}} &= C_g^{-} Q_{g}^{-} + \textrm{H.c.},
\eeq
with
\beq
Q_{g}^{-} & = - \frac{ g_s}{ (4 \pi)^2} \left( \overline{s} \sigma^{\mu \nu} T^A \gamma_5 d \right) G^A_{\mu \nu}.
\eeq
The complete expression for the Wilson coefficient $C_g^{-}$ at the $\mu^{\textrm{SUSY}}$ scale is shown in appendix~\ref{App:WCCg}, where  $(\delta_d)_{13} (\delta_d)_{32}$ terms are discarded for simplicity.
The corresponding loop functions $I (x,y )$, $J (x,y )$, $K (x,y )$, $L (x,y )$, $M_3(x)$, and $M_4(x)$ are defined in appendix~\ref{App:chromo}.

The  chromomagnetic-dipole contribution to $\varepsilon^{\prime}_K /\varepsilon_K$ is \cite{Buras:1999da} 
\beq
\left. \frac{\varepsilon^{\prime}_K}{\varepsilon_K} \right|_{\textrm{chromo}} & = 
\frac{ \omega_{+}}{ |\varepsilon_K^{\textrm{EXP}}|\textrm{Re}A_0^{\textrm{EXP}}}
\left( 1 - \hat{\Omega}_{{\rm eff}} \right)  \frac{11 \sqrt{3}}{ 64 \pi^2} \frac{ M^2_{\pi} M_K^2}{ f_{\pi} (m_s + m_d)} \eta_s B_G \textrm{Im}  C_g^{-},
\eeq
where $f_{\pi}  = (130.2 \pm 1.7 )$ MeV \cite{Patrignani:2016xqp}, and \cite{Cirigliano:2003nn,Cirigliano:2003gt,Buras:2015yba}
\beq
\hat{\Omega}_{{\rm eff}} & =  0.148 \pm 0.080,\\
\eta_{s} &=
 \left[ \frac{\alpha_s (m_b) }{\alpha_s (1.3\textrm{\,GeV})} \right]^{\frac{2}{25}}
 \left[ \frac{\alpha_s (m_t) }{\alpha_s (m_b)} \right]^{\frac{2}{23}} 
 \left[ \frac{\alpha_s (\mu^{\textrm{SUSY}}) }{\alpha_s (m_t)} \right]^{\frac{2}{21}}.
  \eeq
  
According to refs.~\cite{Buras:1999da,Barbieri:1999ax}, the hadronic matrix element for the chromomagnetic-dipole operator into two pions, $B_G$, is enhanced by $ 1/N_c \cdot M_K^2 /M_{\pi}^2$ from the large next-to-leading-order corrections that it receives.
Therefore, the leading order in the chiral quark model, $B_G = 1$, is implausible, and we consider  $ B_G  = 1 \pm 3$ in our analyses.

The other contributions are negligible \cite{Kitahara:2016otd}. 
Note that the sub-leading contributions which come from the gluino-mediated  photon-penguin  and the chargino-mediated $Z$-penguins induced by terms of first order in the expansion of the squark mass matrix, have opposite sign and practically cancel each other  \cite{Kitahara:2016otd}. 

Finally, the SUSY contributions to $\varepsilon^{\prime}_K / \varepsilon_K$ are given as 
\beq
\left( \frac{\varepsilon^{\prime}_K}{\varepsilon_K} \right)^{\textrm{SUSY}}
\simeq 
\left. \frac{\varepsilon^{\prime}_K}{\varepsilon_K} \right|_{\textrm{box}+\textrm{pen}}
 + \left. \frac{\varepsilon^{\prime}_K}{\varepsilon_K} \right|_{\textrm{chromo}}.
\eeq
Note that we discarded the contributions to $\varepsilon^{\prime}_K / \varepsilon_K$ from the heavy Higgs exchanges, although 
they give the strong isospin-violating contribution naturally:
the contribution is enhanced by $\tan^3 \beta$ for only down-type four-fermion scalar operators.
These contributions must be proportional to $m_d m_s$ which cannot be compensated by $\tan^3 \beta$, so that they should be the higher-order contributions for $\varepsilon^{\prime}_K / \varepsilon_K$.

\subsection{$\boldsymbol{\varepsilon_K}$ and $\boldsymbol{\Delta M_K}$}

Although $\varepsilon_K$ is one of the most sensitive quantities to new physics, the SM prediction is still controversial. 
Especially, the leading short-distance contribution to $\varepsilon_K$ in the SM is proportional to $|V_{cb}|^4$ (cf., ref.~\cite{Bailey:2015tba}), 
whose measured values from inclusive semileptonic $B$ decays ($\overline{B} \to X_c \ell^- \overline{\nu}$) 
and from exclusive decays ($\overline{B} \to D^{(\ast)} \ell^- \overline{\nu} $ and $\Lambda_b \to \Lambda_c \ell^- \overline{\nu}$) are inconsistent at a $4.1\sigma$ level 
\cite{Amhis:2016xyh, Jang:2017ieg}. A recent discussion about the exclusive  $|V_{cb}|$ is given  in refs.~\cite{Bigi:2017njr, Grinstein:2017nlq, Bernlochner:2017xyx}.

In this paper, for the SM prediction, we use \cite{Endo:2017ums}
\beq
\varepsilon_K^{\rm SM} = \left( 2.12 \pm 0.18 \right) \times 10^{-3},
\eeq
with
\beq
\varepsilon_K = e^{i \varphi_{\varepsilon} } \varepsilon_K^{\textrm{SM}},
\eeq
where $\varphi_{\varepsilon} =\tan ^{-1} (2 \Delta M_K / \Delta \Gamma_K) =  (43.51 \pm 0.05)^{\circ}$ \cite{Patrignani:2016xqp}.
This value and the uncertainty are based on the inclusive $|V_{cb}|$ \cite{Jang:2017ieg},  the Wolfenstein parameters in the angle-only-fit method \cite{Bevan:2013kaa}, and the long-distance contribution obtained by the lattice simulation \cite{Bai:2015nea}.
Combining the measured value in eq.~\eqref{eq:epKEXP}, we impose
\beq
\varepsilon_K^{\rm EXP/SM} =  1.05\pm 0.10\text{(TH)},
\label{eq:epsKSM}
\eeq
on the $\chi^2$ test, 
with 
\beq
 \varepsilon_K^{\textrm{SM}} \to \varepsilon_K^{\textrm{SM}}  + \varepsilon_K^{\textrm{SUSY}}.
\eeq
Note that we also impose $\textrm{Re} (\varepsilon_K) > 0$ from $\textrm{Re} (\varepsilon_K) = ( 1.596 \pm 0.013) \times 10^{-3}$ \cite{Ambrosino:2006ek}.

Within the MSSM, the SUSY contributions to $\varepsilon_K$ are dominated by  gluino box diagrams.
In this paper, however, we will focus on their suppressed region.
The crossed and uncrossed gluino-box diagrams give opposite sign contributions and there is a certain cancellation region \cite{Crivellin:2010ys,Kitahara:2016otd}, and/or simultaneous mixings of $(\delta_d^{LL})$ and $(\delta_d^{RR})$ can also produce the cancellation.
Therefore, we also consider the sub-dominant contributions which come from Wino and Higgsino boxes.
The $|\Delta S | = 2 $ four-quark effective Hamiltonian at the $\mu^{\rm SUSY}$ scale is \cite{Gabbiani:1996hi}
\beq
\mathcal{H}_{\textrm{eff}} = \sum_{i=1}^{5} C_i Q_i + \sum_{i=1}^{3} \tilde{C}_i  \tilde{Q}_i  + \textrm{H.c.},
\eeq
with
\beq
Q_1 &= \left(  \overline{{d}} \gamma_{\mu} P_L s \right) \left(  \overline{{d}} \gamma^{\mu} P_L s \right),~~~Q_2 = \left( \overline{{d}} P_L s \right) \left( \overline{{d}} P_L s \right),~~~
Q_3 = \left( \overline{{d}}_{\alpha} P_L s_{{\beta}} \right) \left( \overline{{d}}_{\beta} P_L s_{{\alpha}} \right),\non
Q_4 &= \left( \overline{{d}}P_L  s \right) \left( \overline{{d}}P_R  s \right),~~~~~~~~~\,
Q_5 = \left( \overline{{d}}_{\alpha}P_L  s_{{\beta}} \right) \left( \overline{{d}}_{\beta}P_R s_{{\alpha}} \right),\non
\tilde{Q}_1 &= \left(  \overline{{d}} \gamma_{\mu} P_R s \right) \left(  \overline{{d}} \gamma^{\mu} P_R s \right), ~~~
\tilde{Q}_2 = \left( \overline{{d}} P_R s \right) \left( \overline{{d}} P_R s \right),~~~
\tilde{Q}_3 = \left( \overline{{d}}_{\alpha}P_R s_{{\beta}} \right) \left( \overline{{d}}_{\beta} P_R s_{{\alpha}} \right).
\eeq
The kaon mixing amplitude $M_{12}^{(K) }$, $\Delta M_K$ and $\varepsilon_K$ are given by
\beq
M_{12}^{(K) } &= \frac{\langle K^0 | \mathcal{H}_{\textrm{eff}}| \overline{K}{}^0 \rangle}{2 M_K}, \label{eq:M12K}\\
\Delta M_K &= 2 \textrm{Re}[ M_{12}^{(K)} ],\\
\varepsilon_{K} &= \kappa_{\varepsilon} \frac{e^{i \varphi_{\varepsilon}}}{\sqrt{2}} \frac{\textrm{Im}[ M_{12}^{(K)}]}{\Delta M_K^{\textrm{EXP}}}= e^{i \varphi_{\varepsilon} }  \varepsilon_K^{\rm SUSY} ,
\eeq 
where 
$\kappa_{\varepsilon} = 0.94 \pm 0.02$ \cite{Buras:2010pza}.
Using the latest lattice result \cite{Garron:2016mva}, for the hadronic matrix elements, we obtain
\beq
\langle K^0 | \vec{Q} (\mu) | \overline{K}{}^0 \rangle= \Bigl( 0.00211, -0.04231, 0.01288, 0.09571, 0.02452  \Bigr) ~\textrm{(GeV})^4, 
\eeq
with $\langle K^0 | \tilde{Q}_{1,2,3} (\mu) | \overline{K}{}^0 \rangle = \langle K^0 | Q_{1,2,3} (\mu) | \overline{K}{}^0 \rangle$, 
where $\mu = 3 $ GeV and we used $m_s(\mu) = (81.64 \pm 1.17)$ MeV and $ m_d (\mu) = (2.997 \pm 0.049)$ MeV \cite{ Garron:2016mva}.

The leading-order QCD RG corrections are given by \cite{Bagger:1997gg}
\beq
C_{1}(\mu ) &= \eta^K_1 C_{1}(\mu^{\rm SUSY}),\\
\begin{pmatrix}
C_{2}(\mu )\\
C_{3}(\mu )\end{pmatrix}
 & = 
 X_{23} \eta^K_{23} X^{-1}_{23}  \begin{pmatrix}
C_{2}(\mu^{\rm SUSY})\\
C_{3}(\mu^{\rm SUSY})\end{pmatrix},\\
\begin{pmatrix}
C_{4}(\mu )\\
C_{5}(\mu )\end{pmatrix}
 & = 
\begin{pmatrix}
(\eta^K_{1})^{-4} & \frac{1}{3} \left[ (\eta^K_{1})^{-4}  - (\eta^K_{1})^{\frac{1}{2}} \right]\\
0 & (\eta^K_{1})^{\frac{1}{2}}
\end{pmatrix}
\begin{pmatrix}
C_{4}(\mu^{\rm SUSY})\\
C_{5}(\mu^{\rm SUSY})\end{pmatrix},
\eeq
with
\beq
\eta^K_1 &= \left[ \frac{\alpha_s (m_b) }{\alpha_s (\mu)} \right]^{\frac{6}{25}} 
 \left[ \frac{\alpha_s (m_t) }{\alpha_s (m_b)} \right]^{\frac{6}{23}} 
 \left[ \frac{\alpha_s (\mu^{\rm SUSY}) }{\alpha_s (m_t)} \right]^{\frac{6}{21}},\\
\eta^K_{23} & =  \begin{pmatrix} (\eta^K_{1})^{\frac{1}{6} \left( 1 -  \sqrt{241} \right)} & 0 \\ 0 &  (\eta^K_{1})^{\frac{1}{6} \left( 1 +  \sqrt{241} \right)} \end{pmatrix},\\
X_{23} &=   \begin{pmatrix}
 \frac{1}{2}\left( -15 -\sqrt{241} \right) & \frac{1}{2} \left( -15 + \sqrt{241} \right) \\
 1 & 1 \end{pmatrix}.
\eeq
The QCD corrections to $\tilde{C}_{1,2,3}$ are the same as 
${C}_{1,2,3}$.

The Wilson coefficients from the $|\Delta S | = 2$ gluino boxes are shown in appendix~\ref{App:WCgluinoDel2} with their corresponding loop functions defined in appendix~\ref{App:gluinoboxcontribution}. In the universal squark mass limit, these results are consistent  with ref.~\cite{Altmannshofer:2009ne}.
Here, the terms proportional to $ \left[ (\mathcal{M}_D^2)_{LR} \right]_{12}$ or  $(\delta_d)_{13} (\delta_d)_{32}$ are discarded for simplicity.

The Wilson coefficients and their corresponding  loop functions for the sub-leading contributions to $\varepsilon_K$ are given in appendix~\ref{App:WCwinoDel2} and \ref{App:winohiggsino}, respectively.


\section{Parameter scan}
\label{sec:scan}
The MSSM parameter scan is performed with the framework \texttt{Ipanema-$\beta$}~\cite{Ipanema} using a GPU of the model GeForce GTX 1080. The samples are a combination of
flat scans plus scans based on genetic algorithms~\cite{IEEE}. 
The cost function used by the genetic algorithm is the likelihood function with the observable constrains. In addition, 
aiming to get a dense population in regions with $\mathcal{B}(K_S^0\rightarrow\mu^+\mu^-)$ significantly different from the SM prediction, specific penalty contributions are added to the total cost function. We also perform specific scans at $\tan\beta \approx 50$ and
$M_A \approx 1.6$ TeV as for those values the chances to get sizable MSSM effects are
larger.

We study three different scenarios (for the ranges of the scanned parameters see table~\ref{tab:AllScenarios}): 
\begin{itemize}
\item Scenario A: A generic scan with universal gaugino masses. No constraint on the Dark Matter relic density is applied in this case, other than the requirement of neutralino Lightest Supersymmetric Particle (LSP).
The LSP is Bino-like in most cases, although some points with Higgsino LSP are also found.
\item Scenario B: A scan motivated by scenarios with Higgsino Dark Matter. In this
scenario, the relic density is mostly function of the LSP mass, which fulfills the
measured density~\cite{Planck} at $m_{\chi_1^0} \approx 1$ TeV~\cite{Costa:2017gup, Bagnaschi:2017tru,Bagnaschi:2015eha, Bagnaschi:2016xfg}. Thus, we perform a scan with $|\mu| = 1$ TeV $< M_1$. We assume universal gaugino masses in this scenario, which 
then implies that $M_3> 4.5$ TeV.
\item Scenario C: A scan motivated by scenarios with Wino Dark Matter, which is possible
in mAMSB or pMSSM, although it is under pressure by $\gamma$-rays and antiprotons data~\cite{Cuoco:2017iax}. In those scenarios, the relic density is mostly function of the LSP mass, 
which fulfills the experimental value~\cite{Planck} at $m_{\chi_1^0} \approx 3$ TeV  \cite{Hisano:2006nn,Bagnaschi:2016xfg}. Thus, we make a scan with 
$M_2 = 3$ TeV $< |\mu|, M_{1,3}$. The Bino mass $M_1$ is set to 5 TeV for simplicity. Since it is only necessary in order to ensure that the LSP is Wino-like, any other value above 3 TeV (such as, e.g., an mAMSB-like relation $M_1\approx 9.7$ TeV) could also be used without changing the obtained results. The 
lightest neutralino and the lightest chargino are nearly degenerate, and radiative corrections
are expected to bring the chargino mass to be $\approx 160$ MeV heavier than the lightest neutralino  \cite{Ibe:2012sx}.
\end{itemize}
For simplicity, in all cases we set to zero the trilinear couplings and the mass insertions 
other than $\left(\delta_{d}^{LL(RR)}\right)_{12}$ and $\left( \delta_{u}^{LL} \right)_{12}$ which is given by the relations
in eq.~\eqref{eq:duLL}, and $\mu$ is treated as a real parameter, with
both signs allowed a priori.

\begin{table}[t!]
\begin{center}
\begin{tabular}{cccc}
Parameter & Scenario A & Scenario B & Scenario C \\
\hline 
$\tilde{m}_{Q}$ & [2, 10] & [2, 10] & [4, 10]  \\
$\tilde{m}_{Q}^2/\tilde{m}_{d}^2$ & [0.25, 4] & [0.25, 4] & [0.25, 4] \\
$M_3$ & [2, 10] & [4.5, 15] & [4, 15] \\
$\tan\beta$ & [10, 50] & [10, 50] & [10, 50] \\
$M_A$ & [1, 2] & [1, 2] & [1, 2] \\
$|\mu|$ & [1, 10] & $1$ & [5, 20]\\
$M_1$ & $\frac{\alpha_1(\mu^{SUSY})}{\alpha_3(\mu^{SUSY})}M_3$ & $\frac{\alpha_1(\mu^{SUSY})}{\alpha_3(\mu^{SUSY})}M_3$ & 5 \\
$M_2$ & $\frac{\alpha_2(\mu^{SUSY})}{\alpha_3(\mu^{SUSY})}M_3$ & $\frac{\alpha_2(\mu^{SUSY})}{\alpha_3(\mu^{SUSY})}M_3$ & 3 \\
$B_G$ & [-2, 4] & [-2, 4] & [-2, 4]\\
$\textrm{Re}\left[(\delta_{d}^{LL(RR)})_{12}\right]$& [-0.2, 0.2] & [-0.2, 0.2] & [-0.2, 0.2] \\
$\textrm{Im}\left[(\delta_{d}^{LL(RR)})_{12}\right]$& [-0.2, 0.2] & [-0.2, 0.2] & [-0.2, 0.2] \\
\hline 
\end{tabular}
\caption{\label{tab:AllScenarios}Scan ranges for scenario A, B (motivated by Higgsino Dark Matter) and C (motivated by Wino Dark Matter). All masses are in TeV.
The nuisance parameter $B_G$ appears in the chromomagnetic-dipole contribution to $\varepsilon^{\prime}_K/\varepsilon_K$.}
\end{center}
\end{table}

We also perform studies at the MFV limit, using RG equations induced MIs in CMSSM.
As expected, no significant effect is found in this case.

For the squark masses, we use $\tilde{m}_{Q} = \tilde{m}_{u} \neq \tilde{m}_{d}$.
This set up is motivated by the SUSY $\textrm{SU}(5)$ grand unified theory, where
$Q$ and $U$-squark are contained in $\mathbf{10}$ representation matter multiplet  
while $D$-squark is in  $\mathbf{\overline{5}}$ representation one.
In general, their soft-SUSY breaking masses are different and depend on couplings between the matter 
multiplets and the SUSY breaking spurion field.

\section{Results}
\label{sec:results}
In the following, we show the main results of our scans.
The points with $\chi^2<12.5$, corresponding to $95\%$ C.L. for six degrees of freedom, are considered experimentally viable.
The number of degrees of freedom has been calculated as the number
of observables, not counting the nuisance parameter $B_G$,
the rigid bound on the $\tan\beta$:$M_A$ plane, and $\Delta M_K$, which
are not Gaussian distributed. Therefore, the $\chi^2$ requirement
corresponds to a $95\%$ C.L. or tighter. Similar plots are obtained
if one uses a looser bound on the absolute $\chi^2$ accompanied with a $\Delta\chi^2 <5.99$ across the plane being plotted. 
Due to the large theory uncertainty, $\mathcal{B}(K_L^0\rightarrow\mu^+\mu^-)$ can go up to $\approx 1\times 10^{-8}$ at 2$\sigma$ level. Values slightly
above that limit can still be allowed if they reduce the $\chi^2$ contribution in other observables. The allowed regions are separated
by the sign of $A^\mu_{L\gamma\gamma}$ in eq.~\eqref{eq:ALgg}. We also show results for
$A_{CP}$, which could be experimentally accessed by means of a tagged analysis.

\subsection{Effects from $\left(  \delta_{d}^{LL(RR)} \right)_{12}$ separately}

\begin{figure}[t!]
\centering
\includegraphics[width=0.49\textwidth]{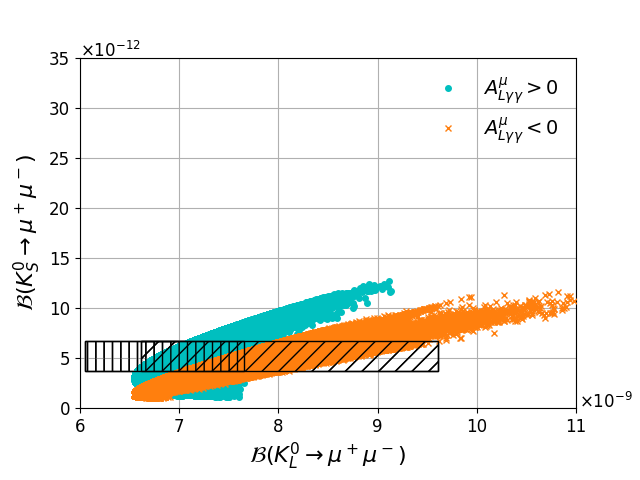}
\includegraphics[width=0.49\textwidth]{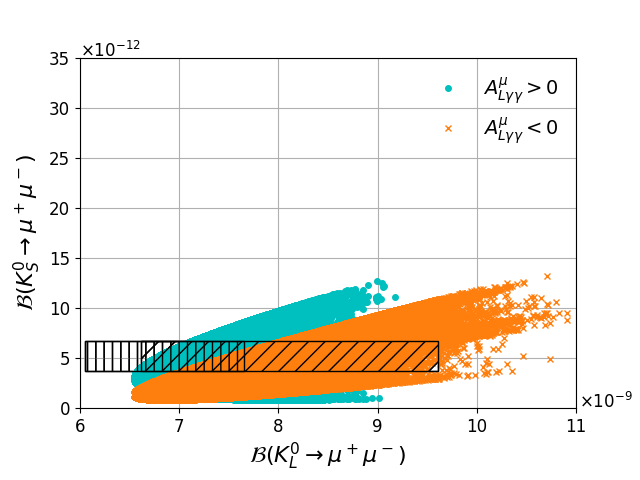}\\
\includegraphics[width=0.49\textwidth]{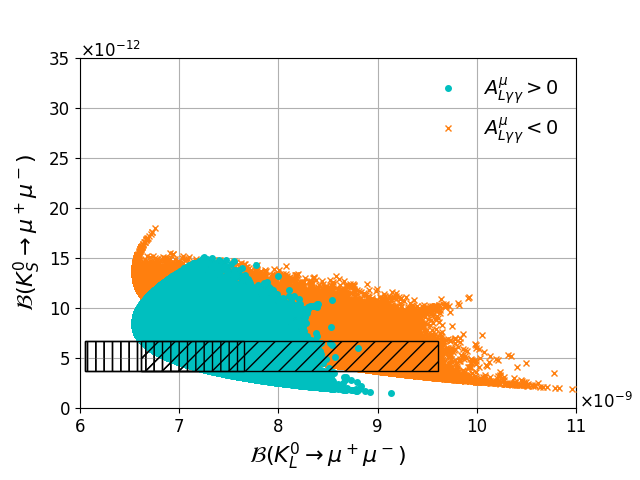}
\includegraphics[width=0.49\textwidth]{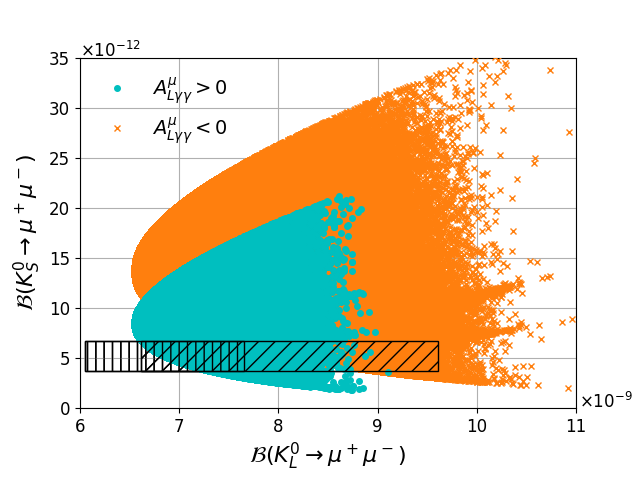}
\caption{\label{fig:BR_SCA} Scenario A $\mathcal{B}(K_S^0\rightarrow\mu^+\mu^-)$ vs $\mathcal{B}(K_L^0\rightarrow\mu^+\mu^-)$ for $\left(\delta_{d}^{LL}\right)_{12}\neq 0$ and $(M_3\cdot\mu)>0$ (upper left), $\left(\delta_{d}^{LL}\right)_{12}\neq 0$ and $(M_3\cdot\mu)<0$ (upper right), $\left(\delta_{d}^{RR}\right)_{12}\neq 0$ and $(M_3\cdot\mu)>0$ (lower left), and $\left(\delta_{d}^{RR}\right)_{12}\neq 0$ and $(M_3\cdot\mu)<0$ (lower right). The cyan dots correspond to $A^\mu_{L\gamma \gamma} > 0$ and the orange crosses to $A^\mu_{L\gamma \gamma} < 0$. The vertically hatched area corresponds to the SM prediction for $A^\mu_{L\gamma \gamma} > 0$ and the inclined hatched area corresponds to the SM prediction for $A^\mu_{L \gamma \gamma} < 0$.}
\end{figure}

\begin{figure}[h!]
\centering
\includegraphics[width=0.49\textwidth]{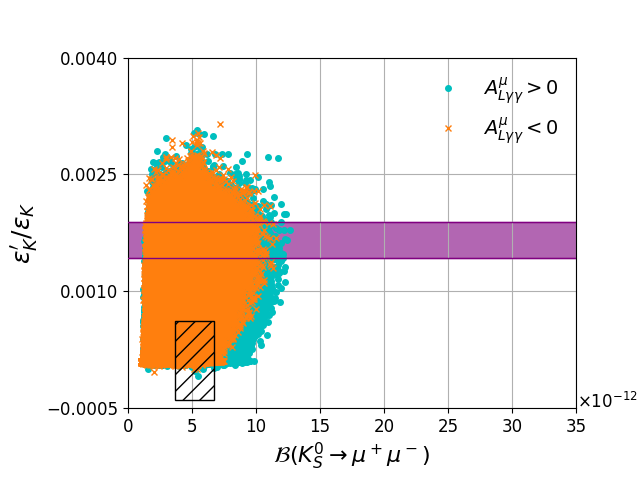}
\includegraphics[width=0.49\textwidth]{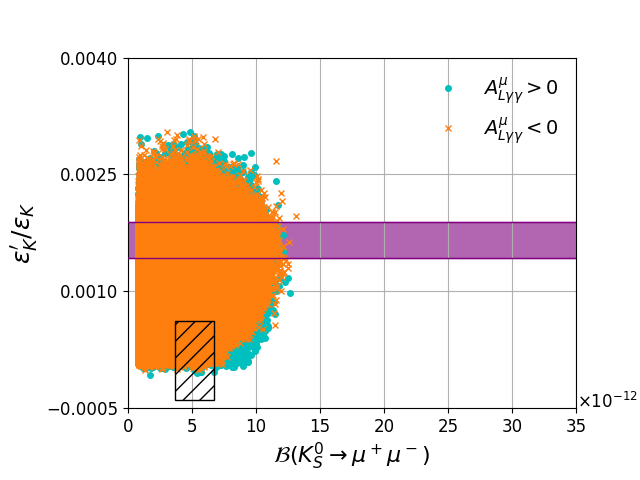}\\
\includegraphics[width=0.49\textwidth]{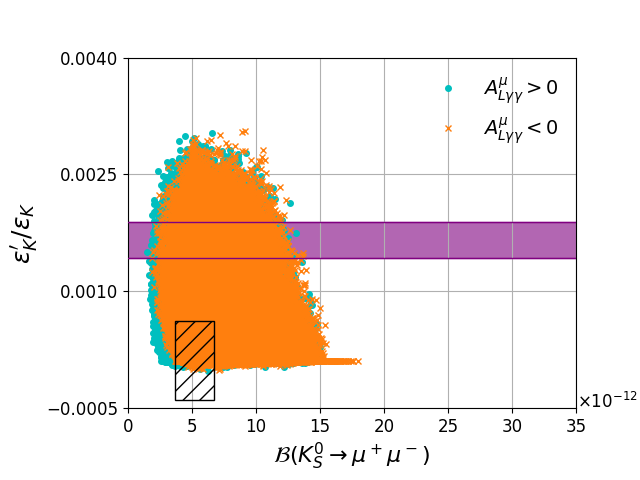}
\includegraphics[width=0.49\textwidth]{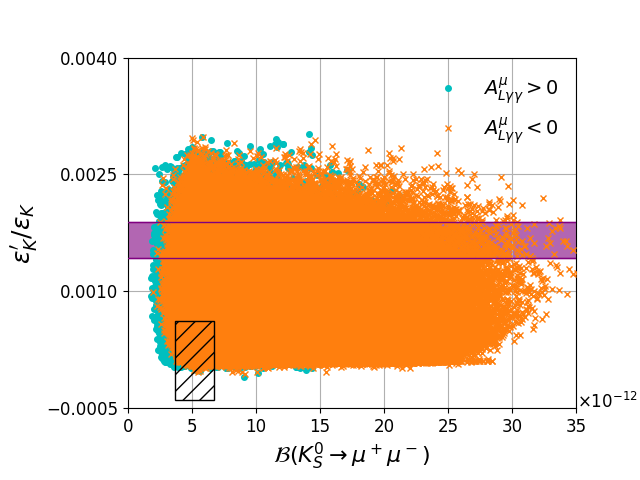}
\caption{\label{fig:eps_SCA} Scenario A $\frac{\varepsilon ^\prime_K}{\varepsilon_K}$ vs $\mathcal{B}(K_S^0\rightarrow\mu^+\mu^-)$ for $\left(\delta_{d}^{LL}\right)_{12}\neq 0$ and $(M_3\cdot\mu)>0$ (upper left), $\left(\delta_{d}^{LL}\right)_{12}\neq 0$ and $(M_3\cdot\mu)<0$ (upper right), $\left(\delta_{d}^{RR}\right)_{12}\neq 0$ and $(M_3\cdot\mu)>0$ (lower left), and $\left(\delta_{d}^{RR}\right)_{12}\neq 0$ and $(M_3\cdot\mu)<0$ (lower right). The cyan dots correspond to $A^\mu_{L\gamma \gamma} > 0$ and the orange crosses to $A^\mu_{L\gamma \gamma} < 0$. The deep purple band corresponds to the experimental results and the hatched area to the SM prediction.}
\end{figure}

\begin{figure}[h!]
\centering
\includegraphics[width=0.49\textwidth]{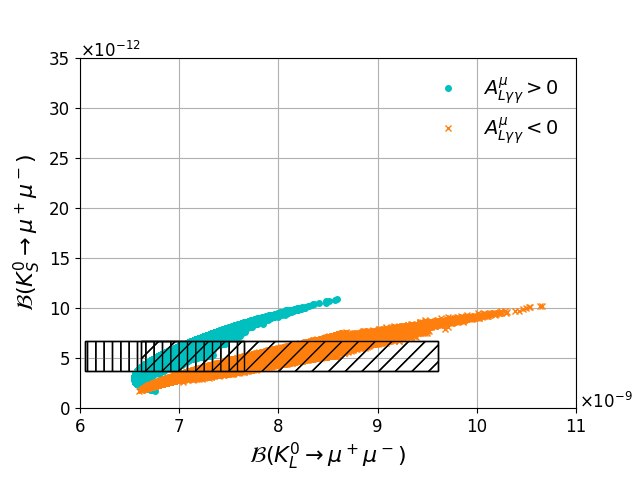}
\includegraphics[width=0.49\textwidth]{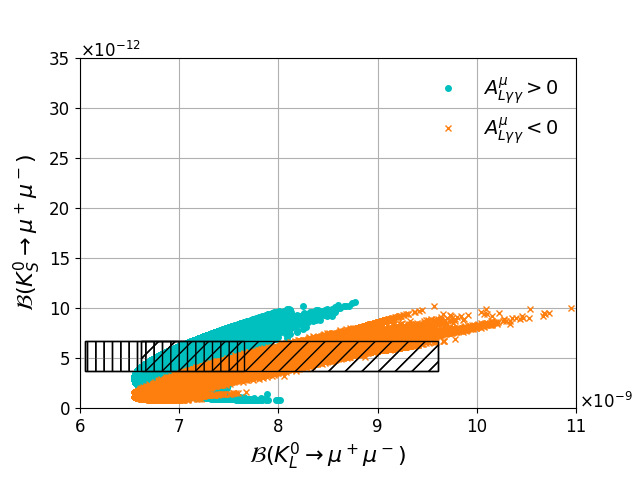}\\
\includegraphics[width=0.49\textwidth]{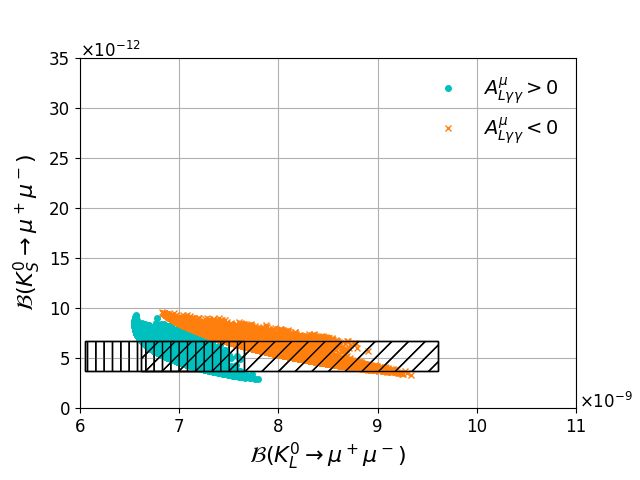}
\includegraphics[width=0.49\textwidth]{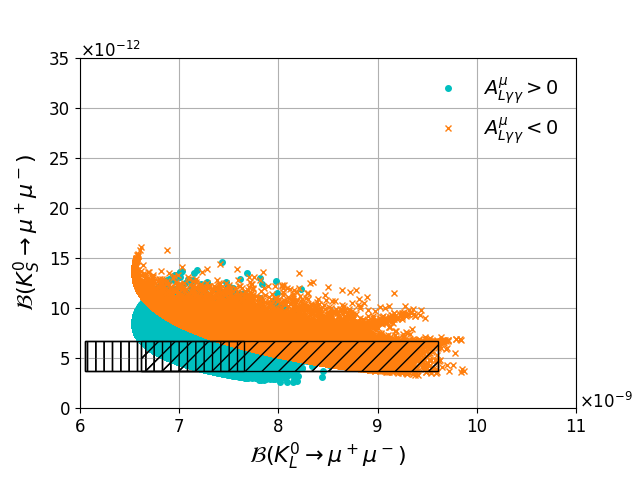}
\caption{\label{fig:BR_SCB} Scenario B, motivated by Higgsino Dark Matter with universal gaugino masses, $\mathcal{B}(K_S^0\rightarrow\mu^+\mu^-)$ vs $\mathcal{B}(K_L^0\rightarrow\mu^+\mu^-)$ for $\left( \delta_{d}^{LL} \right)_{12}\neq 0$ and $(M_3\cdot\mu)>0$ (upper left), $\left(\delta_{d}^{LL}\right)_{12}\neq 0$ and $(M_3\cdot\mu)<0$ (upper right), $\left(\delta_{d}^{RR}\right)_{12}\neq 0$ and $(M_3\cdot\mu)>0$ (lower left), and $\left( \delta_{d}^{RR} \right)_{12}\neq 0$ and $(M_3\cdot\mu)<0$ (lower right). The cyan dots correspond to $A^\mu_{L\gamma \gamma} > 0$ and the orange crosses to $A^\mu_{L\gamma \gamma} < 0$. The vertically hatched area corresponds to the SM prediction for $A^\mu_{L\gamma \gamma} > 0$ and the inclined hatched area corresponds to the SM prediction for $A^\mu_{L\gamma \gamma} < 0$.}
\end{figure}

\begin{figure}[h!]
\centering
\includegraphics[width=0.49\textwidth]{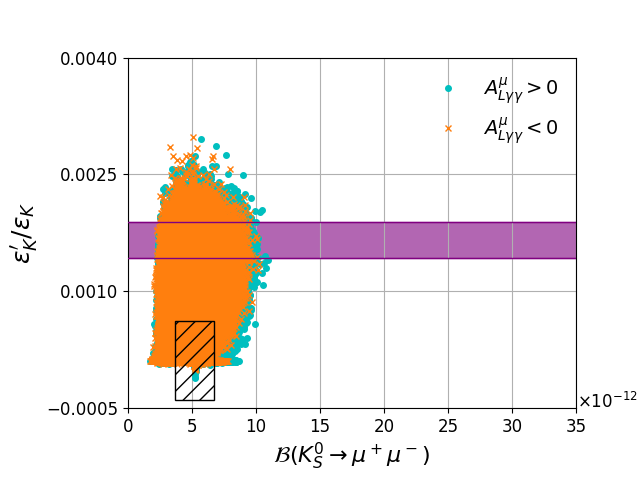}
\includegraphics[width=0.49\textwidth]{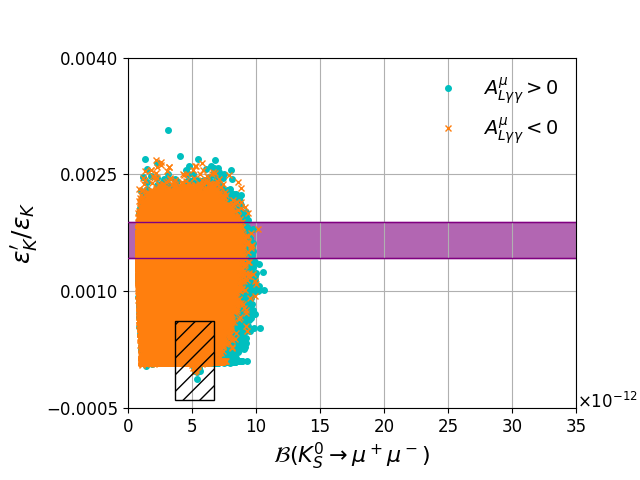}\\
\includegraphics[width=0.49\textwidth]{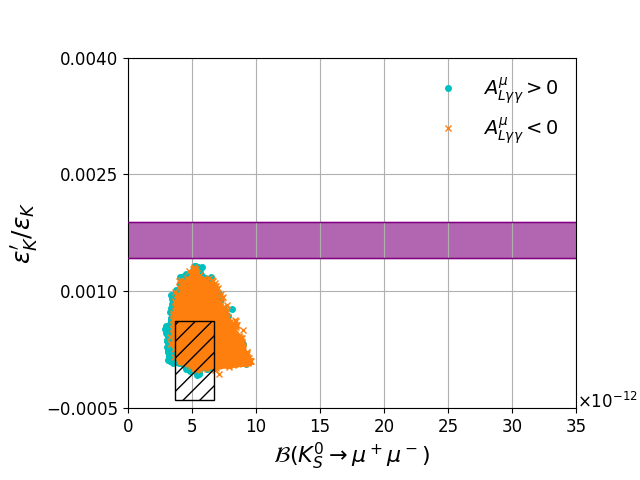}
\includegraphics[width=0.49\textwidth]{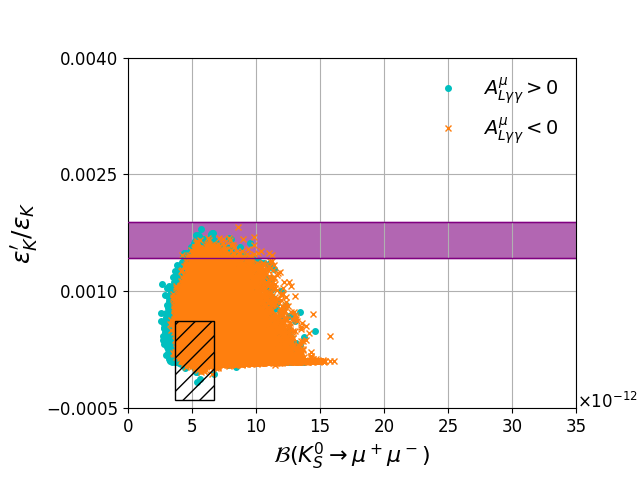}
\caption{\label{fig:eps_SCB} Scenario B, motivated by Higgsino Dark Matter and universal gaugino masses, $\frac{\varepsilon ^\prime_K}{\varepsilon_K}$ vs $\mathcal{B}(K_S^0\rightarrow\mu^+\mu^-)$ for $\left( \delta_{d}^{LL} \right)_{12}\neq 0$ and $(M_3\cdot\mu)>0$ (upper left), $\left(\delta_{d}^{LL}\right)_{12}\neq 0$ and $(M_3\cdot\mu)<0$ (upper right), $\left(\delta_{d}^{RR}\right)_{12}\neq 0$ and $(M_3\cdot\mu)>0$ (lower left), and $\left( \delta_{d}^{RR} \right)_{12}\neq 0$ and $(M_3\cdot\mu)<0$ (lower right). The cyan dots correspond to $A^\mu_{L\gamma \gamma} < 0$ and the orange crosses to $A^\mu_{L\gamma \gamma} > 0$. The deep purple band corresponds to the experimental results and the hatched area to the SM prediction.}
\end{figure}

\begin{figure}[h!]
\centering
\includegraphics[width=0.49\textwidth]{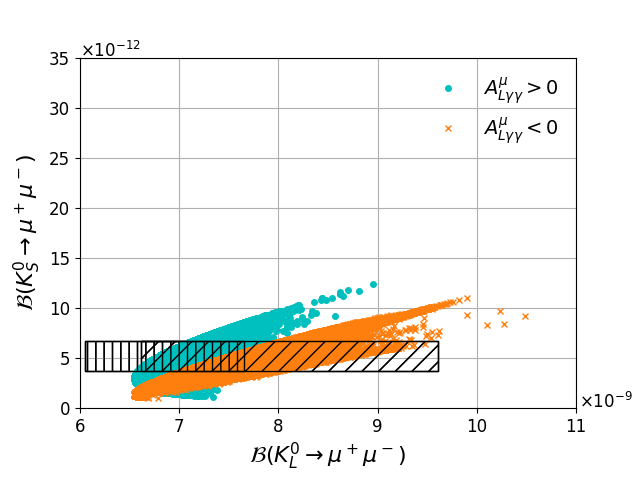}
\includegraphics[width=0.49\textwidth]{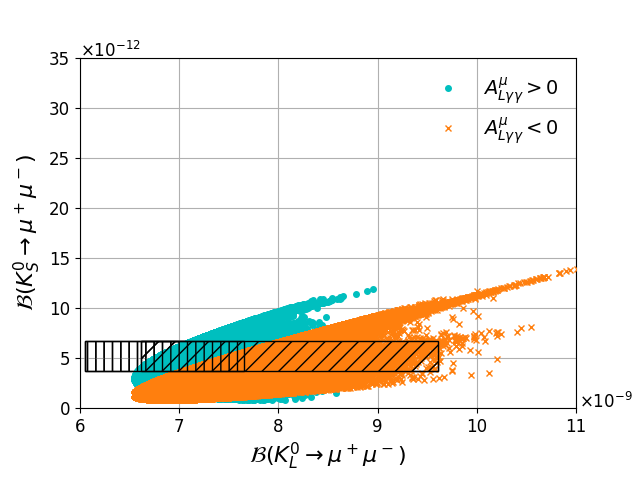}\\
\includegraphics[width=0.49\textwidth]{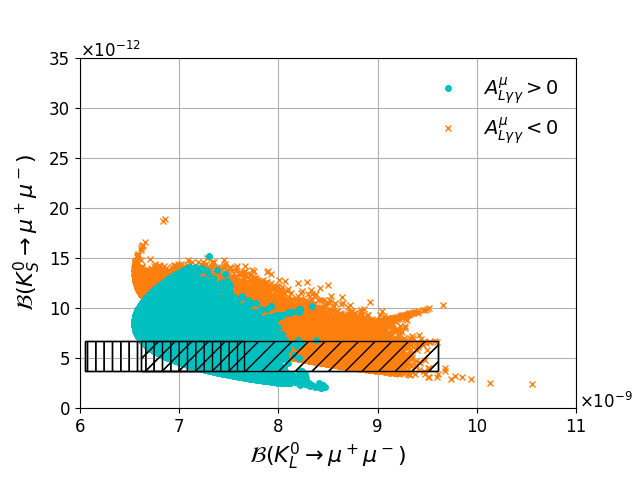}
\includegraphics[width=0.49\textwidth]{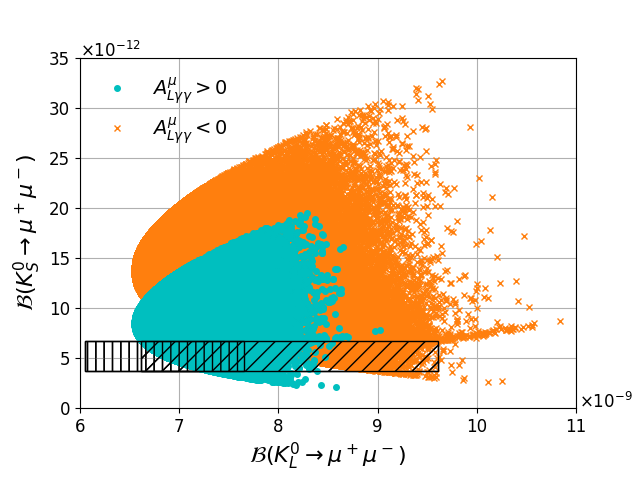}
\caption{\label{fig:BR_SCC} Scenario C (motivated by Wino Dark Matter) $\mathcal{B}(K_S^0\rightarrow\mu^+\mu^-)$ vs $\mathcal{B}(K_L^0\rightarrow\mu^+\mu^-)$ for $\left(\delta_{d}^{LL}\right)_{12}\neq 0$ and $(M_3\cdot\mu)>0$ (upper left), $\left(\delta_{d}^{LL}\right)_{12}\neq 0$ and $(M_3\cdot\mu)<0$ (upper right), $\left(\delta_{d}^{RR}\right)_{12}\neq 0$ and $(M_3\cdot\mu)>0$ (lower left), and $\left(\delta_{d}^{RR}\right)_{12}\neq 0$ and $(M_3\cdot\mu)<0$ (lower right). The cyan dots correspond to $A^\mu_{L\gamma \gamma} > 0$ and the orange crosses to $A^\mu_{L\gamma \gamma} < 0$. The vertically hatched area corresponds to the SM prediction for $A^\mu_{L\gamma \gamma} > 0$ and the inclined hatched area corresponds to the SM prediction for $A^\mu_{L\gamma \gamma} < 0$.}
\end{figure}

\begin{figure}[h!]
\centering
\includegraphics[width=0.49\textwidth]{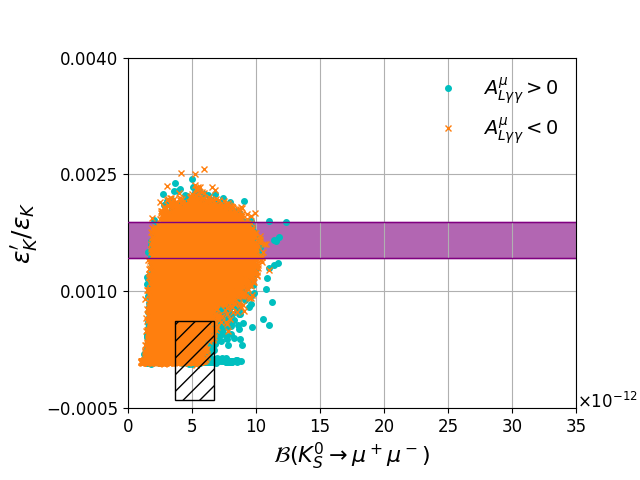}
\includegraphics[width=0.49\textwidth]{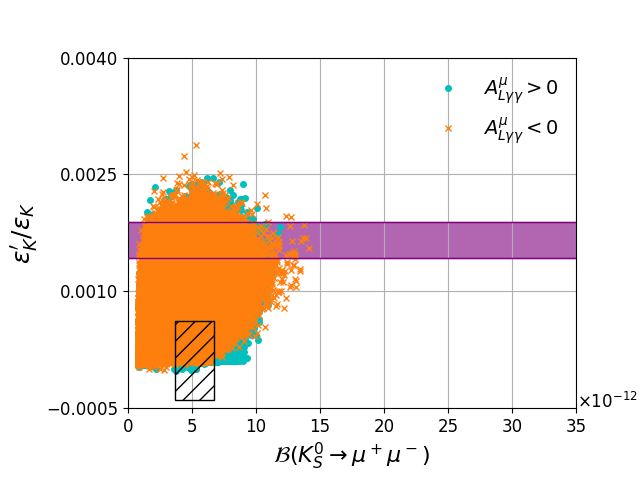}\\
\includegraphics[width=0.49\textwidth]{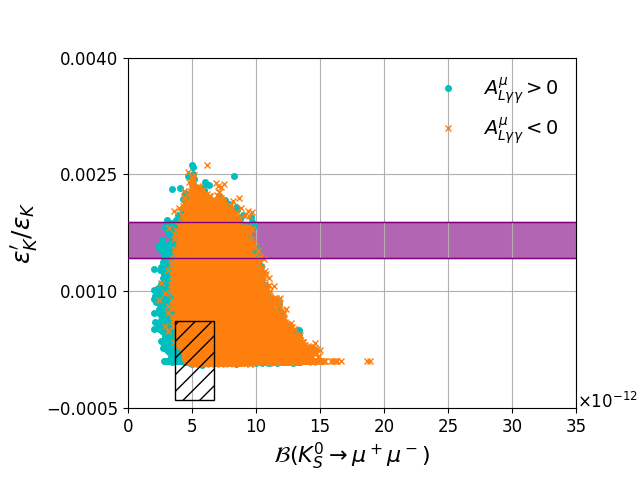}
\includegraphics[width=0.49\textwidth]{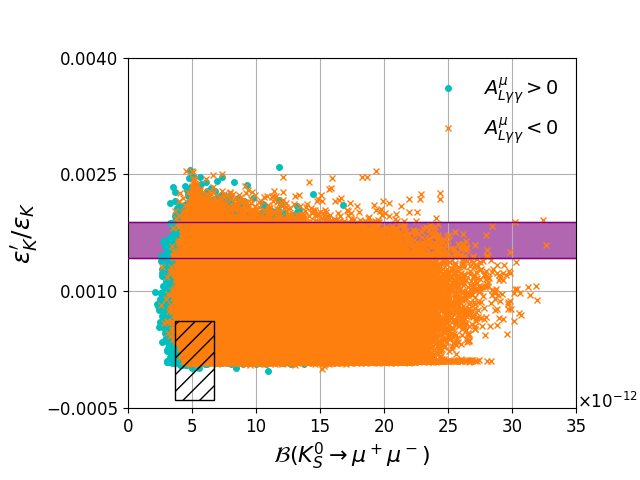}
\caption{\label{fig:eps_SCC} Scenario C, motivated by Wino Dark Matter, $\frac{\varepsilon ^\prime_K}{\varepsilon_K}$ vs $\mathcal{B}(K_S^0\rightarrow\mu^+\mu^-)$ for $\left(\delta_{d}^{LL}\right)_{12}\neq 0$ and $(M_3\cdot\mu)>0$ (upper left), $\left(\delta_{d}^{LL}\right)_{12}\neq 0$ and $(M_3\cdot\mu)<0$ (upper right), $\left(\delta_{d}^{RR}\right)_{12}\neq 0$ and $(M_3\cdot\mu)>0$ (lower left), and $\left(\delta_{d}^{RR}\right)_{12}\neq 0$ and $(M_3\cdot\mu)<0$ (lower right). The cyan dots correspond to $A^\mu_{L\gamma \gamma} > 0$ and the orange crosses to $A^\mu_{L\gamma \gamma} < 0$. The deep purple band corresponds to the experimental results and the hatched area to the SM prediction.}
\end{figure}

We first study separately the effects of pure left-handed or
pure right-handed MIs, to study the regions of the MSSM parameter space in which either $LL$ MIs or $RR$ MIs dominate\footnote{As 
an example, MFV models the $LL$ MIs can become non-zero after RGE,
which does not happen for $RR$ MIs.}.
The obtained scatter plots for $\mathcal{B}(K_L^0\rightarrow\mu^+\mu^-)$  vs $\mathcal{B}(K_S^0\rightarrow\mu^+\mu^-)$ and $\mathcal{B}(K_S^0\rightarrow\mu^+\mu^-)$ vs $\varepsilon^\prime_K /\varepsilon_K$
are shown in figure~\ref{fig:BR_SCA} and figure~\ref{fig:eps_SCA} for 
Scenario A, figure~\ref{fig:BR_SCB} and figure~\ref{fig:eps_SCB} for
Scenario B, and figure~\ref{fig:BR_SCC} and figure~\ref{fig:eps_SCC} for
Scenario C. 
The points in the planes correspond to predictions from different
values of the input parameters. 
One should note that in such cases, the SUSY contributions to 
$\varepsilon_K$ can be suppressed naturally in a heavy gluino region ($M_3  \gtrsim 1.5 \tilde{m}_{Q}$) \cite{Crivellin:2010ys,Kitahara:2016otd}.


In Scenario A (see figure~\ref{fig:BR_SCA}) and Scenario C (see figure~\ref{fig:BR_SCC}),
we can see that the $95\%$ C.L. allowed regions for $\mathcal{B}(K_S^0\rightarrow\mu^+\mu^-)$
in light of the constraints listed in table~\ref{tab:Observables} are approximately
$[0.78,14]\times 10^{-12}$ for $LL$-only contributions, and $[1.5,35]\times 10^{-12}$  for $RR$-only 
contributions, without any need of fine-tuning the parameters to avoid constraints from $\mathcal{B}(K_L^0\rightarrow\mu^+\mu^-)$. The MSSM contributions are similar
for $RR$ and $LL$, and the differences on the allowed ranges for $\mathcal{B}(K_S^0\rightarrow\mu^+\mu^-)$ arise from  the interference with the SM amplitudes in 
$K_{S(L)}^0\rightarrow\mu^+\mu^-$,
which are shown in section~\ref{sec:KSmumu}.
The allowed regions for scenarios A and C are very similar to each other, although marginally larger on A.
It can also be seen that,  in Scenario B (see figure~\ref{fig:BR_SCB}) the maximum departure
of $\mathcal{B}(K_S^0\rightarrow\mu^+\mu^-)$ from the SM is smaller than in the other scenarios, since
$C_{S,P} \propto \mu$ and $\mu$ is small relative to squark and gluino masses. 
In the contributions to $(\varepsilon^{\prime}_K / \varepsilon_K)^{\rm SUSY}$, 
the chromomagnetic-dipole contribution can be significant in both $LL$-only and $RR$-only cases when $\mu \tan \beta $ and $B_G$ have large values, while the box contributions can be significant only via $LL$ MIs \cite{Kitahara:2016otd}. 
Note that the penguin contributions to  $(\varepsilon^{\prime}_K / \varepsilon_K)^{\rm SUSY}$ 
are neglected in our parameter scan.

\begin{figure}
\centering
\includegraphics[width=0.49\textwidth]{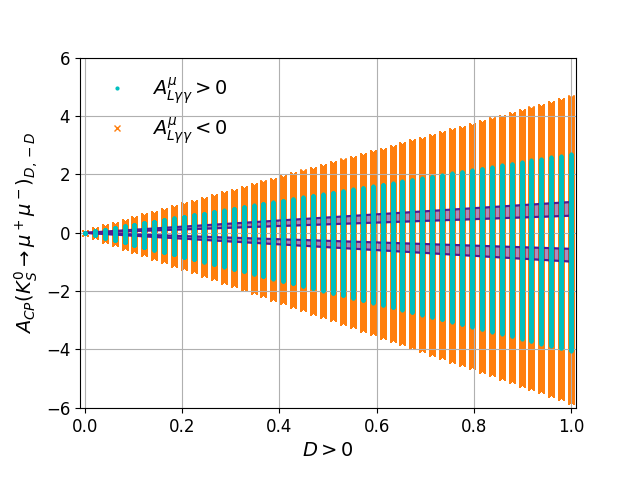}
\includegraphics[width=0.49\textwidth]{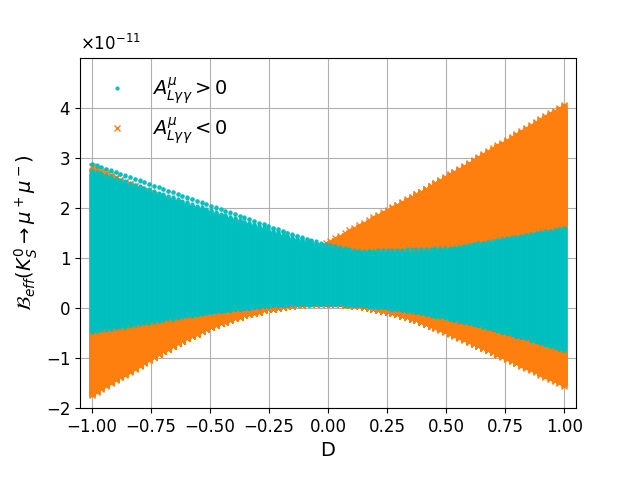}
 	\caption{\label{fig:Acp_BReff} Scenario A, $\left( \delta^{LL}_d \right)_{12} \neq 0$ and $(M_3 \cdot \mu) < 0$. Plots of $A_{CP}(K^0_S \rightarrow \mu^+ \mu^-)$ vs $D$ (left) for the case $D = -D^\prime$ ($D > 0$) where the cyan dots correspond to $A^\mu_{L\gamma \gamma} > 0$, the orange crosses to $A^\mu_{L\gamma \gamma} < 0$, and the deep purple bands correspond to the SM predictions in eq.~\eqref{sq:ACPSM}.  ${\cal B}(K^0_S \rightarrow \mu^+ \mu^-)_{{\rm eff}}$ vs $D$ (right).} 
\end{figure}

\begin{figure}
\centering
\includegraphics[width=0.49\textwidth]{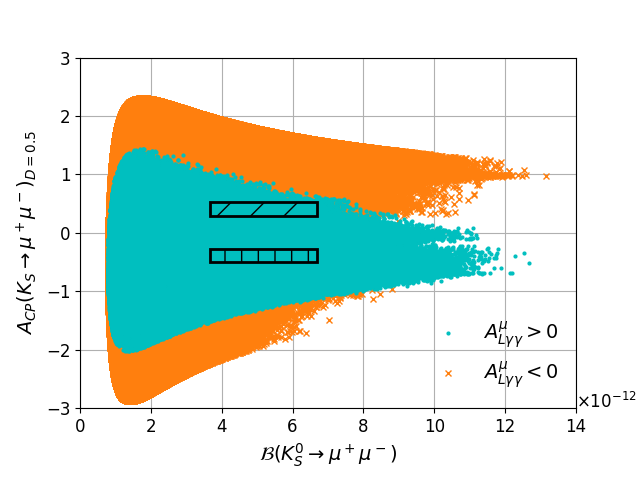}
\includegraphics[width=0.49\textwidth]{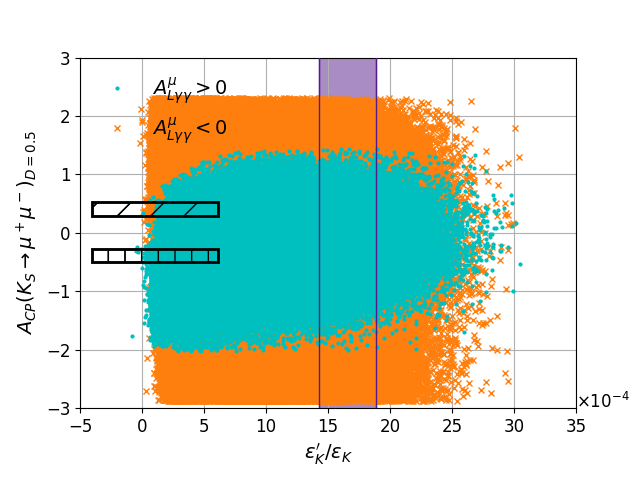}
\includegraphics[width=0.49\textwidth]{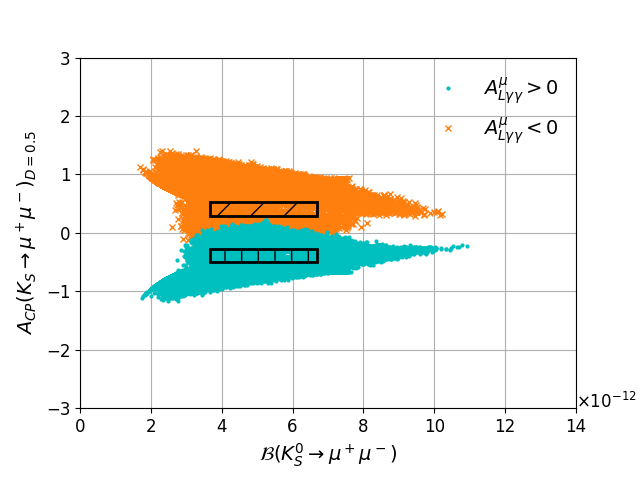}
\includegraphics[width=0.49\textwidth]{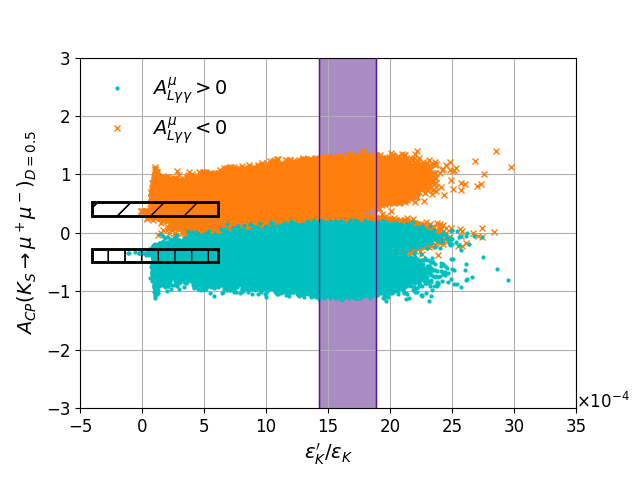}
 	\caption{\label{fig:Acp_SCB}$A_{CP}$ vs $\mathcal{B}(K^0_S \rightarrow \mu^+ \mu^-)$ (left) and vs $\varepsilon^\prime_K/\varepsilon_K$ (right). The top panels correspond to Scenario A, $\left( \delta_d^{LL} \right)_{12} \neq 0$ and $(M_3 \cdot \mu) < 0$. The bottom panels correspond to Scenario B, $\left( \delta^{LL}_d \right)_{12} \neq 0$ and $(M_3 \cdot \mu) > 0$. The plots are done for $D = -D^\prime = 0.5$ . The cyan dots correspond to $A^\mu_{L\gamma \gamma} > 0$ and the orange crosses to $A^\mu_{L\gamma \gamma} < 0$. The deep purple bands correspond to the experimental value of $\varepsilon^\prime_K/\varepsilon_K$, the vertically hatched areas correspond to the SM prediction for $A^\mu_{L\gamma \gamma} > 0$ and the inclined hatched areas to the SM prediction for $A^\mu_{L\gamma \gamma} < 0$ .} 
\end{figure}

The effective branching fraction and $CP$ asymmetry are shown in figure~\ref{fig:Acp_BReff} for Scenario A.Note that the negative value of ${\cal B}(K^0_S \rightarrow \mu^+ \mu^-)_{{\rm eff}}$ is compensated in data by inclusion of the background events from $K^0_L \rightarrow \mu^+ \mu^-$, so that the overal $K^0\rightarrow \mu^+\mu^-$ is always positive. Correlation patterns of $A_{CP}$ with other observables can be seen in figure~\ref{fig:Acp_SCB},
where we choose $D' = -D$ and $D = 0.5$ for simplicity . We find that $CP$ asymmetries can be up to $\approx 6$ (at $D = 1$), approximately eight times
bigger than in the SM. The largest effects are found in left-handed scenarios.

\clearpage
\newpage

\subsection{Floating $LL$ and $RR$ MIs simultaneously}

A priori, one possibility to avoid the constraint from $\mathcal{B}(K_L^0\rightarrow\mu^+\mu^-)$ is 
to allow simultaneously for non-zero $LL$ and $RR$ mass insertions. This way both $C_{S(P)}$ and $\tilde{C}_{S(P)}$ are non zero
and eqs.~\eqref{eq:FF1}--\eqref{eq:FF2} do not hold. One can then find regions in which the MSSM contributions to  $\mathcal{B}(K_S^0\rightarrow\mu^+\mu^-)$ do not alter $\mathcal{B}(K_L^0\rightarrow\mu^+\mu^-)$ significantly.

\begin{figure}
\centering
\includegraphics[width=0.49\textwidth]{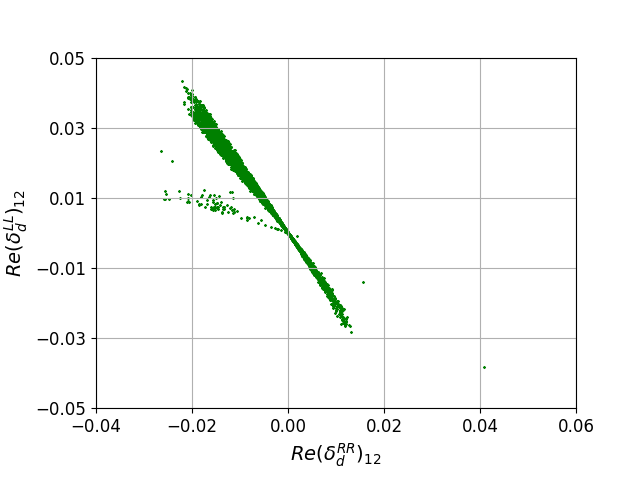}
\includegraphics[width=0.49\textwidth]{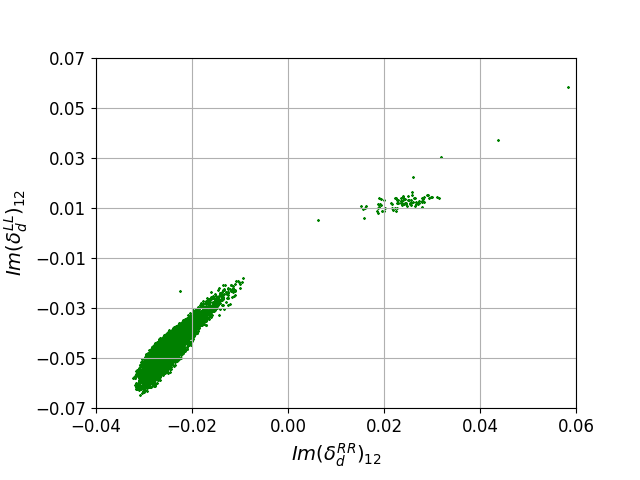}
\includegraphics[width=0.49\textwidth]{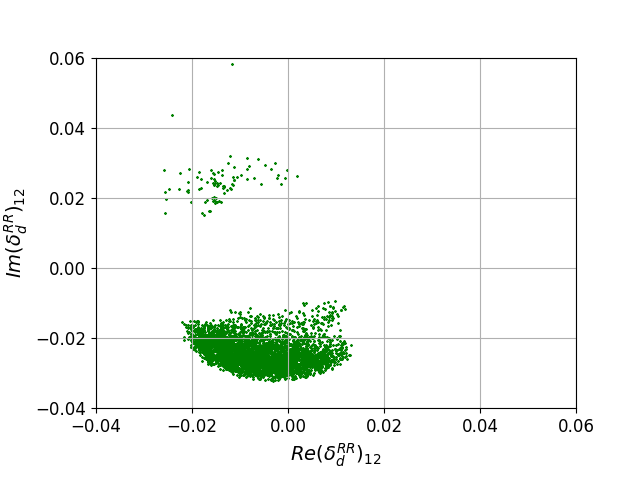}
\includegraphics[width=0.49\textwidth]{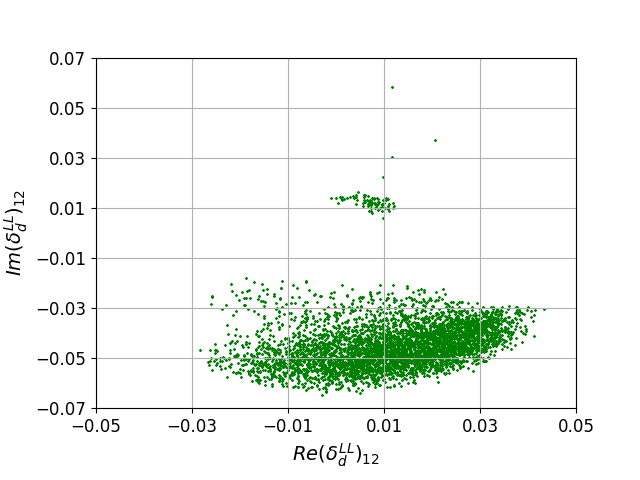}
 	\caption{\label{fig:strips} Scatter plots of the real (upper left) and the imaginary (upper right) parts of the mass insertions $\left(\delta_d^{RR}\right)_{12}$ and $\left(\delta_d^{LL}\right)_{12}$ for $\mathcal{B}(K_S^0\rightarrow\mu^+\mu^-) > 2\times10^{-10}$, of the real vs imaginary $\left(\delta_d^{RR}\right)_{12}$ (lower  left) and of the real vs imaginary $\left(\delta_d^{LL}\right)_{12}$ (lower right). All points in the plane pass the experimental constraints defined in section~\ref{sec:formalism}. The up-type MI $(\delta_u^{LL})_{12}$ is given by eq.~\eqref{eq:duLL}. The plots correspond to Scenario C, with a sample of 4378 points with $\mathcal{B}(K_S^0\rightarrow\mu^+\mu^-) > 2\times10^{-10}$ and $\chi^2<12.5$, produced after 6M generations of $200k$ points each. The pattern observed in Scenario A is very similar.}
\end{figure}

\begin{figure}
\centering
\includegraphics[width=0.49\textwidth]{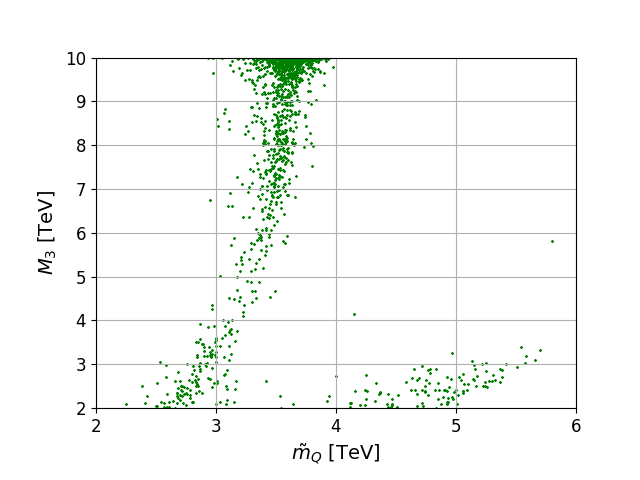}
\includegraphics[width=0.49\textwidth]{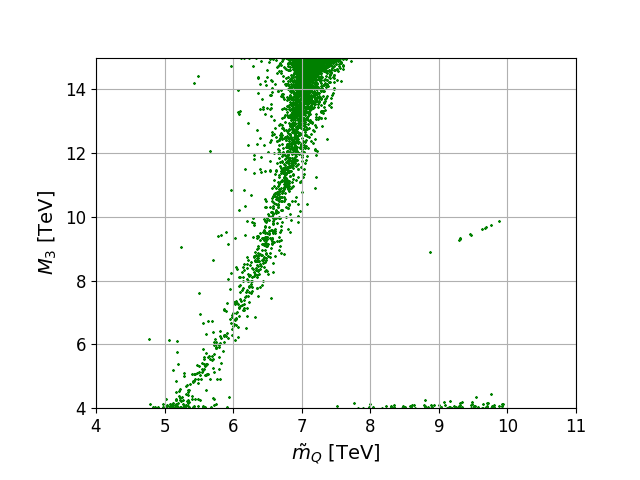}
\caption{\label{fig:msq_vs_mg} Scatter plot of the squark and gluino masses for $\mathcal{B}(K_S^0\rightarrow\mu^+\mu^-) > 2\times 10^{-10}$ taking into account the constraints defined in section~\ref{sec:formalism}. Left: Scenario A, Right: Scenario C. The $\chi^2$ cut in Scenario A has been relaxed to $14$ to increase the density of points.}
\end{figure}

For instance, if one chooses
\beq
\textrm{Re} \left[ \left(\delta_{d}^{LL}\right)_{12} \right] = - \textrm{Re} \left[ \left(\delta_{d}^{RR}\right)_{12} \right], \quad 
\textrm{Im} \left[ \left(\delta_{d}^{LL}\right)_{12} \right] = \textrm{Im} \left[ \left(\delta_{d}^{RR}\right)_{12} \right],
\label{eq:strips}
\eeq
then the SUSY contributions to $\mathcal{B}(K_L^0\rightarrow\mu^+\mu^-)$ are canceled, while the SUSY contributions to $\mathcal{B}(K_S^0\rightarrow\mu^+\mu^-)$ are maximized (see eqs.~\eqref{eq:brKSLmm}--\eqref{eq:brKSLmmend}).
However, it is known that in those cases the
bounds from $\Delta M_K$ and $\varepsilon_K$ are very stringent. 
Using genetic algorithms with cost functions that target large values of $\mathcal{B}(K_S^0\rightarrow\mu^+\mu^-)$, we find fine-tuned regions with $\mathcal{B}(K_S^0\rightarrow\mu^+\mu^-) > 10^{-10}$, or even at the level of
the current experimental bound of $8\times 10 ^{-10}$ at $90\%$ C.L.~\cite{LHCb:KsMuMu}, which
are consistent with all our constraints.
These points are located along very narrow strips in the $\left(\delta_{d}^{LL}\right)_{12}$ vs $\left(\delta_{d}^{RR}\right)_{12}$ planes, as shown in figure~\ref{fig:strips}. The figure corresponds to Scenario C as it is the one with higher density of points at large 
values of $\mathcal{B}(K_S^0\rightarrow\mu^+\mu^-)$ and the pattern observed in Scenario
A is nearly identical.
A particularly favorable region corresponds to $|(\delta_d^{LL})_{12}| \approx 2|(\delta_d^{RR})_{12}| \sim 0.03$
and $arg\left[(\delta_d^{LL})_{12}\right] \approx -arg\left[(\delta_d^{RR})_{12}\right] + \pi$, which is in the vicinity of eq.~\eqref{eq:strips}, and
with $\delta_u^{LL}$ given by the symmetry relation of eq.~\eqref{eq:duLL}. 
They also favor narrow regions in the squark vs gluino masses planes as shown in figure~\ref{fig:msq_vs_mg}. We checked that the values close to the experimental upper bound
can still be obtained even if the constraint on $\Delta M_K$ is significantly tightened.

We note that the authors in ref.~\cite{Jang:2017ieg} provide a SM prediction for $\varepsilon_K$ less consistent with data than the one we used. That prediction is obtained using $|V_{cb}|$ from exclusive decays. If we use that value instead of eq.~\eqref{eq:epsKSM},
\beq
\varepsilon_K^{\rm EXP/SM} =  1.41\pm 0.16\text{(TH)},
\eeq
then we
can accommodate more easily $LL$ and $RR$ MIs of similar sizes, and fine-tuned regions
with  $\mathcal{B}(K_S^0\rightarrow\mu^+\mu^-) > 10^{-10}$ are found with higher chances. The shapes of the strips in the mass insertion planes do not change substantially.

\clearpage

\subsection{Non degenerate Higgs masses}
The results so far have been obtained in the MSSM framework, in which $|C_S| \approx |C_P|$. This
is due to the mass degeneracy $M_H\approx M_A$. In models in which such degeneracy can be broken, the constraint that $\mathcal{B}(K_L^0\rightarrow\mu^+\mu^-)$
imposes to $\mathcal{B}(K_S^0\rightarrow\mu^+\mu^-)$ relaxes the more those two masses differ.
This degeneracy is broken in MSSM at low values of $M_A$, and requiring $\tan\beta$ to be small to avoid constraints
from $\tan\beta : M_A$ planes from LHC. Those regions are more difficult to study, since it would require a detailed specification
of the MSSM and test it against bounds of the Higgs sector. The mass degeneracy is also broken in extensions such as NMSSM.
According to our scans, on those cases one could, in principle, reach values of
$\mathcal{B}(K_S^0\rightarrow\mu^+\mu^-) > 10^{-10}$ for mass differences of ${\cal O}(33\%)$ or larger without fine-tuning the MIs.

\section{Conclusions}
\label{sec:conclusions}
We explored MSSM contribution to $\mathcal{B}(K_S^0\rightarrow\mu^+\mu^-)$ for non-zero $(\delta_{d}^{LL})_{12}$ and $(\delta_{d}^{RR})_{12}$  mass insertions,
motivated by the experimental value of $\varepsilon'_K/\varepsilon_K$, and in the large $\tan{\beta}$ regime. The expressions for the relevant MSSM amplitudes have been provided.
We find that MSSM contributions to $\mathcal{B}(K_S^0\rightarrow\mu^+\mu^-)$ 
can surpass the SM contributions [$\mathcal{B}(K_S^0\rightarrow\mu^+\mu^-)^{\textrm{SM}} = 5.18\times 10^{-12}$] 
by up to
a factor of seven (see figure~\ref{fig:BR_SCA}), reaching the level of $3.5\times10^{-11}$ even for large SUSY masses, with no conflict with existing experimental data, and are detectable by LHCb. 
This is also the case even if $\varepsilon'_K/\varepsilon_K$ turns out to be SM-like as predicted by refs.~\cite{Pallante:2001he,Hambye:2003cy,Mullor}. Figures of correlations between $\mathcal{B}(K_S^0\rightarrow\mu^+\mu^-)$ and other observables have been provided
for different regions of the MSSM parameter space, and can be used to understand which scenarios are more or less favoured, depending on the experimental outcomes.
The $3.5\times10^{-11}$ bound is due to the combined effect of $\Delta M_K,\varepsilon_K$ , and $K^0_L\rightarrow\mu^+\mu^-$ constraints. Such bound is not rigid, and
fine-tuned regions can bring the branching fraction above the 
$10^{-10}$ level, even up to the current experimental bound; the largest
deviations from SM are found at $|(\delta_d^{LL})_{12}| \approx 2|(\delta_d^{RR})_{12}| \sim 0.03$
and $arg\left[(\delta_d^{LL})_{12}\right] \approx -arg\left[(\delta_d^{RR})_{12}\right] + \pi$ for large squark and
gluino masses. We also find that the $CP$ asymmetry of $K^0\rightarrow\mu^+\mu^-$ can be
significantly modified by MSSM contributions, being up to eight times bigger than the SM prediction in the pure LL case. Finally, we 
remind that, for simplicity, we have restricted our study to the main contributions in the large $\tan\beta$ regime. Discarded terms could,
in principle, provide even more flexibility to the allowed regions.

\acknowledgments 

We would like to thank A. Crivellin,  G. Isidori, T. Kuwahara, D. Mueller, and K.A. Olive for useful discussions. 
The research activity of IGFAE/USC members is partially funded by
ERC-StG-639068 and partially by XuntaGal.
G.~D.~was supported
in part by MIUR under Project No. 2015P5SBHT
(PRIN 2015) and by the INFN research initiative ENP.
K. Y. was supported by Grant-in-Aid for Scientific research from the Ministry of Education, Science, Sports, and Culture (MEXT), Japan, No. 16H06492.

\appendix

\section{Wilson coefficients}

\subsection{$\boldsymbol{|\Delta S | = 1}$ gluino box contribution} 
\label{App:WCgluino}

The Wilson coefficients of the gluino box contributions to $\varepsilon^{\prime}_K/\varepsilon_K$ are
\beq
C_1^{q} &= \frac{ (\alpha_s)^2 }{ 2 \sqrt{2} G_F M_3^2} \left( \delta_d^{LL}\right)_{12} \left[ \frac{1}{18} f \left( x^Q_3, x^q_3  \right)    - \frac{5}{18}  g \left(   x^Q_3, x^q_3 \right)  \right],\non
C_2^{q} &= \frac{ (\alpha_s)^2 }{ 2 \sqrt{2} G_F M_3^2} \left( \delta_d^{LL}\right)_{12} \left[ \frac{7}{6} f \left(  x^Q_3, x^q_3 \right)    + \frac{1}{6}  g \left(  x^Q_3, x^q_3 \right)  \right],\non
C_3^{q} &= \frac{ (\alpha_s)^2 }{ 2 \sqrt{2} G_F M_3^2} \left( \delta_d^{LL}\right)_{12} \left[ - \frac{5}{9} f \left(  x^Q_3, x^Q_3 \right)    + \frac{1}{36}  g \left(   x^Q_3, x^Q_3 \right)  \right],\non
C_4^{q} &= \frac{ (\alpha_s)^2 }{ 2 \sqrt{2} G_F M_3^2} \left( \delta_d^{LL}\right)_{12} \left[ \frac{1}{3} f \left(   x^Q_3, x^Q_3\right)    + \frac{7}{12}  g \left(   x^Q_3, x^Q_3 \right)  \right],\non
\tilde{C}_1^{q} &= \frac{ (\alpha_s)^2 }{ 2 \sqrt{2} G_F M_3^2} \left( \delta_d^{RR}\right)_{12} \left[ \frac{1}{18} f \left( x^d_3, x^Q_3 \right)    - \frac{5}{18}  g \left(  x^d_3, x^Q_3 \right)  \right],\non
\tilde{C}_2^{q} &= \frac{ (\alpha_s)^2 }{ 2 \sqrt{2} G_F M_3^2} \left( \delta_d^{RR}\right)_{12} \left[ \frac{7}{6} f \left(  x^d_3, x^Q_3 \right)    + \frac{1}{6}  g \left( x^d_3, x^Q_3  \right)  \right],\non
\tilde{C}_3^{q} &= \frac{ (\alpha_s)^2 }{ 2 \sqrt{2} G_F M_3^2} \left( \delta_d^{RR}\right)_{12} \left[ - \frac{5}{9} f \left( x^d_3, x^q_3 \right)    + \frac{1}{36}  g \left(  x^d_3, x^q_3\right)  \right],\non
\tilde{C}_4^{q} &= \frac{ (\alpha_s)^2 }{ 2 \sqrt{2} G_F M_3^2} \left( \delta_d^{RR}\right)_{12} \left[ \frac{1}{3} f \left( x^d_3, x^q_3\right)    + \frac{7}{12}  g \left( x^d_3, x^q_3\right)  \right],
\eeq
where $q$ runs $u$ and $d$, and $x^Q_3 = \tilde{m}_{Q}^2 / M_3^2$ and $x^q_3 = \tilde{m}_{q}^2 / M_3^2$.

\subsection{$\boldsymbol{|\Delta S | = 1}$ chargino-mediated  $\boldsymbol{Z}$-penguin contribution}
\label{App:WCchargino}

The Wilson coefficients of the chargino-mediated $Z$-penguin  are 
\beq
C_1^{u} & = - \frac{(\alpha_2)^2  \sin^2 \theta_W }{ 12 \sqrt{2} G_F M_W^2}\frac{ \left[ (\mathcal{M}_U^2)_{LR}\right]^{\ast}_{2 3 } \left[(\mathcal{M}_U^2)_{LR}\right]_{ 1 3} }{M_2^4} l \left(  x^Q_2, x^u_2 \right),\non
C_1^{d} & = \frac{(\alpha_2)^2  \sin^2 \theta_W }{ 24 \sqrt{2} G_F M_W^2}\frac{ \left[ (\mathcal{M}_U^2)_{LR}\right]^{\ast}_{2 3 } \left[(\mathcal{M}_U^2)_{LR}\right]_{ 1 3} }{M_2^4} l \left( x^Q_2, x^u_2 \right),\non
C_3^{u} & = \frac{(\alpha_2)^2 }{ 16 \sqrt{2} G_F M_W^2}\left( 1 - \frac{4}{3}\sin^2 \theta_W \right)\frac{ \left[(\mathcal{M}_U^2)_{LR}\right]^{\ast}_{2 3 } \left[(\mathcal{M}_U^2)_{LR}\right]_{ 1 3} }{M_2^4} l \left( x^Q_2, x^u_2 \right),\non
C_3^{d} & = -  \frac{(\alpha_2)^2 }{ 16 \sqrt{2} G_F M_W^2}\left( 1 - \frac{2}{3} \sin^2 \theta_W \right)\frac{ \left[(\mathcal{M}_U^2)_{LR}\right]^{\ast}_{2 3 } \left[(\mathcal{M}_U^2)_{LR}\right]_{ 1 3} }{M_2^4} l \left( x^Q_2, x^u_2\right),\non
C_{2,4}^{q} &= \tilde{C}_{1,2,3,4}^{q}  = 0.
\eeq

\subsection{$\boldsymbol{|\Delta S | = 1}$ chromomagnetic dipole contribution}
\label{App:WCCg}

The Wilson coefficients of the chromomagnetic dipole contributions to $\varepsilon^{\prime}_K/\varepsilon_K$ are
\beq
C_g^{-} &= 
  \frac{\alpha_s \pi }{3}  \frac{ \tilde{m}_{Q}^2 \mu  m_s}{M_3^5 } 
 \left( \delta_d^{LL} \right)_{12} \frac{ \tan \beta}{  1 + \epsilon_g \tan \beta      }
  \left[   I  \left(  x^Q_3, x^d_3  \right) +   9 J  \left(   x^Q_3, x^d_3  \right)\right]\non  
& - \frac{\alpha_s \pi }{3}  \frac{ \tilde{m}_{d}^2 \mu  m_s}{M_3^5 } 
 \left( \delta_d^{RR} \right)_{12} \frac{ \tan \beta}{  1 + \epsilon_g \tan \beta      }
  \left[   I  \left( x^d_3, x^Q_3    \right)  + 9 J  \left(   x^d_3, x^Q_3 \right) \right] \non
  & + \frac{\alpha_s \pi }{3} \frac{   \left[(\mathcal{M}_D^2)_{LR}\right]_{12}  - \left[(\mathcal{M}_D^2)_{LR}\right]^{\ast}_{2 1}  }{M_3^3}
\left[   K \left( x^Q_3, x^d_3 \right)  
+ 9 L \left( x^Q_3, x^d_3  \right)\right]\non
&-  \frac{\alpha_s \pi }{3}  \frac{m_s}{ \tilde{m}_{Q}^2}  \left( \delta_d^{LL} \right)_{12}  
  \left[   M_3  \left(  x^3_Q\right) + 9 M_4 \left(x^3_Q \right)   \right]\non
  & + \frac{\alpha_s \pi }{3}  \frac{m_s}{\tilde{m}_{d}^2}
 \left( \delta_d^{RR} \right)_{12}
  \left[   M_3  \left( x^3_d \right) +  9 M_4 \left( x^3_d \right) \right].
\label{eq:Cgminus}
\eeq

\subsection{$\boldsymbol{|\Delta S | = 2}$ gluino box contribution}
\label{App:WCgluinoDel2}

The Wilson coefficients of the gluino box contributions to $\varepsilon_K$  are
\beq
C_1 &=   - \frac{(\alpha_s)^2}{ \tilde{m}_{Q}^2}  \left[ \left( \delta^{LL}_d \right)_{21} \right]^2 g^{(1)}_1 \left(  x^3_Q \right),  
 \\
C_4 &=   - \frac{(\alpha_s)^2}{ M_3^2}  \left[ \left( \delta^{LL}_d \right)_{21}  \left( \delta^{RR}_d \right)_{21}  \right] g^{(1)}_4 \left(x^3_Q ,  x^3_d \right),
\\
C_5 &\simeq    - \frac{(\alpha_s)^2}{ M_3^2}  \left[ \left( \delta^{LL}_d \right)_{21}  \left( \delta^{RR}_d \right)_{21}  \right] g^{(1)}_5 \left( x^3_Q ,  x^3_d  \right),
\\
\tilde{C}_1 &=  - \frac{(\alpha_s)^2}{ \tilde{m}_{d}^2}  \left[ \left( \delta^{RR}_d \right)_{21} \right]^2 g^{(1)}_1 \left( x^3_d\right), 
\\
C_2  &= C_3 =  \tilde{C}_2 = \tilde{C}_3 =  0.
\eeq

\subsection{Sub-leading contributions to $\boldsymbol{\varepsilon_K}$}
\label{App:WCwinoDel2}

The Wilson coefficients of the Wino and Higgsino contributions are
\beq
C_1 & = - \frac{\alpha_s \alpha_2 }{ 6 \tilde{m}_{Q}^2 }   \left[ \left( \delta^{LL}_d \right)_{21} \right]^2  g^{(1)}_{\tilde{g}\tilde{w}} \left( x^3_Q, x^2_Q   \right)  
%
 - \frac{(\alpha_2)^2 }{ 8 \tilde{m}_{Q}^2 } \left[ \left( \delta^{LL}_d \right)_{21} \right]^2  g^{(1)}_{\tilde{w}} \left( x^2_Q  \right) \non
 & - \frac{ (\alpha_2)^2}{8 \tilde{m}_{u}^2 }  \left( V_{ts} V_{td}^{\ast} \right)^2  \frac{m_t^4}{ M_W^4}    f_1 \left( x^{\mu}_u \right) ,\\
\tilde{C}_3 & = - \frac{ (\alpha_2)^2}{ 8 } \left( V_{ts} V_{td}^{\ast} \right)^2  \frac{m_s ^2 \tan^2 \beta}{(1 + \epsilon_g \tan \beta)^2}\frac{m_t^4}{M_W^4} \frac{\mu^2 A_t^2}{ \tilde{m}_{Q}^4 \tilde{m}_{u}^4} f_3\left( x^{\mu}_Q , x^{\mu}_u \right) ,\\
C_2 & = C_3 = C_4  =  C_5 =\tilde{C}_1 = \tilde{C}_2  =  0.
\eeq

Note that a $\tan^4 \beta$ enhanced contribution to $\varepsilon_K $ comes from the exchange of neutral Higgses, which is discarded because of $\left( \delta_d \right)_{23} \left( \delta_d \right)_{31} = 0$ in our analyses.
For the Wilson coefficient, we obtain 
\beq
C_2 & \simeq \tilde{C}_2  \simeq 0 ,\\
C_4 & \simeq   - \frac{ 8 (\alpha_s)^2 \alpha_2 }{ 9 \pi} \frac{m_b^2}{M_W^2} \frac{\tan^4 \beta}{ (1 + \epsilon_g \tan \beta)^2 [1 + (\epsilon_g  + \epsilon_Y y_t^2 )\tan \beta]^2 } \frac{ \mu^2 M_3^2}{ M_A^2 \tilde{m}_{Q}^2 \tilde{m}_{d}^2}\non
& \times \left[ \left( \delta^{LL}_d \right)_{23} \left( \delta^{LL}_d \right)_{31}\left( \delta^{RR}_d \right)_{23} \left( \delta^{RR}_d \right)_{31} \right] H\left( x^3_Q, x^d_Q \right) H\left( x^3_d, x^Q_d  \right)  ,\\
C_1 & = C_3 = C_5=\tilde{C}_1=\tilde{C}_3 = 0 ,
\eeq
where the approximation in eq.~\eqref{eq:appMA} is used, and the loop function $H(x,y)$ is given in eq.~\eqref{eq:Hloop}.
Note that the $CP$-even and $CP$-odd Higgs contributions to $C_2$ ($\tilde{C}_2$) are canceled out by each other.

\section{Loop functions}

\subsection{$\boldsymbol{K^0 \rightarrow \mu^+ \mu^-}$}
\label{App:loopKmm}

The loop functions $l(x,y)$, $F(x,y)$, $G(x,y)$, and $H(x,y)$ are given by 
\beq
l (x,y) &= - \frac{ \left[ x^2 + (x - 2) y \right] x \ln x }{(x-1)^2 (x-y)^3} + \frac{ \left[ y^2 + (y - 2) x \right] y \ln y }{(y-1)^2 (x - y)^3 } - \frac{x + y - 2 x y}{(x-1) (y-1) (x - y)^2},
\label{LoopL}\\
F (x, y ) &= \frac{ x \ln x}{(x - 1)(x - y)} + \frac{ y \ln y}{ (y-1) (y - x)}, \\
G (x,y) & = \frac{ x \ln x}{(x - 1)^2 (x - y)} + \frac{ y \ln y}{ (y-1)^2 (y - x)} + \frac{1 }{(x - 1)(y-1)},\\
H (x,y) & = \frac{ x \ln x}{ (x-1 )^2 (x - y)^2 } + \frac{ (x + x y - 2 y^2) \ln y}{ (y-1)^3 ( x - y)^2 } - \frac{ 2 x -y -1}{(x - 1) (y-1)^2 ( x - y)}, 
\label{eq:Hloop}
\eeq
where $l (1,1)  = - 1/12$,  $F(1,1) = 1/2$, $G(1,1) = - 1/6$, and $H(1,1 ) = 1/12 $.

\subsection{$\boldsymbol{\varepsilon^\prime _K/ \varepsilon_K}$}

\subsubsection{$\boldsymbol{|\Delta S| = 1}$ gluino box contributions}
\label{App:gluinobox}

The loop functions $f(x,y) $ and $g(x,y)$ \cite{Kagan:1999iq} are 
\beq
f (x, y) & = 
\frac{x[ 2 x^2 - (x +1) y ] \ln x}{(x - 1)^3 ( x - y)^2}
- \frac{ x y \ln y}{(y - 1)^2 (x - y)^2} + 
\frac{x (x + 1 - 2 y )}{(x-1)^2 ( y-1)(x-y)}, \\
g (x, y) & = 
- \frac{x^2 [ x (x + 1) - 2 y] \ln x}{(x - 1)^3 ( x - y)^2}
+ \frac{ x y^2 \ln y}{(y - 1)^2 (x - y)^2} + 
\frac{x [ - 2 x + (x + 1) y]}{(x-1)^2 ( y-1)(x-y)}, 
\eeq
which lead to
\beq
f (x, x) & = 
- \frac{ 1 + 4 x - 5 x^2 + 2 x (2 + x) \ln x} { 2 (  x- 1)^4} = \frac{1}{x} B_2 \left(\frac{1}{x} \right),\\
g (x, x) & = 
\frac{ x \left[ 5 -4 x - x^2 + 2 (1 + 2 x) \ln x \right]}{2 ( x-1)^4} = - \frac{4}{x} B_1 \left( \frac{1}{x} \right).
\eeq
The loop functions $B_{1,2} (x)$ are consistent with ref.~\cite{Gabbiani:1996hi}  for the universal squark masses case.

\subsubsection{Chromomagnetic-dipole operator}
\label{App:chromo}

The loop functions $I (x,y )$, $J (x,y )$, $K (x,y )$, $L (x,y )$, $M_3(x)$, and $M_4(x)$ are given by
\beq
I (x,y ) & = \frac{ (3 x^2 - y - 2 x y) \ln x}{ (x-1)^4 (x-y)^2} - \frac{y \ln y}{(y-1)^3 (x-y)^2} \non
& ~~~ + 
\frac{  -2 + (-5 + x) x + 9 y + (2 + x)x y - (5 + x) y^2}{2 (x - y) (x-1)^3 (y-1)^2},\\
J (x,y) & =   -\frac{x  [ (1 + 2 x) x  - (2 + x) y] \ln x }{ (  x-1)^4 (x - y)^2} 
 + \frac{ y^2 \ln y}{ (y-1)^3 (x-y)^2}  \non
 & ~~~ + \frac{   (5 + x) x - 3  y (1 + x)^2  + (1 + 5 x) y^2  }{2 (x -1)^3 (y -1 )^2 (x - y) }  ,\\
K(x,y) & = \frac{ x \ln x}{ (x-y) (x-1)^3} + \frac{ y \ln y}{( y- x) (y-1)^3} + \frac{x y  + x + y -3  }{2 (x-1)^2 (y-1)^2},\\
L(x,y) & = - \frac{x^2 \ln x}{(x-y ) (x-1)^3} - \frac{y^2 \ln y}{(y-x) (y-1)^3 } + \frac{1 + x + y - 3 x y}{2 (x-1)^2 (y-1)^2},\\
 M_3 (x) & = \frac{ -1 + 9 x + 9 x^2 - 17 x^3 + 6 x^2 (3 + x) \ln x}{12 (  x-1)^5},\\
M_4 (x) & = \frac{ -1 - 9 x + 9 x^2 + x^3 - 6 x (1 + x) \ln x } {6 ( x-1)^5},
\eeq
which lead to
\beq
K(x,x) & = \frac{-5 + 4 x + x^2 - 2 (1 + 2 x) \ln x}{2 ( x-1)^4} = \frac{1}{x^2} M_1 \left( \frac{1}{x} \right),\\
L(x,x) & = \frac{ 1 + 4 x - 5 x^2 + 2 x (2 + x) \ln x} {2 (  x-1)^4} = - \frac{1}{x} B_2 \left( \frac{1}{x} \right).
\eeq
The above $M_{1,3,4} (x)$ are consistent with  ref.~\cite{Gabbiani:1996hi} in the universal squark masses case.\footnote{
We found that in eq.\,(14) of ref.~\cite{Gabbiani:1996hi}, $M_2 (x) = - x B_2 (x)$ should be replaced by $M_2 (x) = - B_2 (x) /x$, which has been pointed out in ref.~\cite{Harnik:2002vs}.}

\subsection{$\boldsymbol{\varepsilon_K}$}
\subsubsection{$\boldsymbol{|\Delta S| = 2}$ gluino box contributions}
\label{App:gluinoboxcontribution}

The loop functions $g^{(1)}_1 (x) $, $g^{(1)}_4 (x, y )$, and $g^{(1)}_5 (x, y )$ are given by
\beq
g^{(1)}_1 (x) & = - \frac{ 11 + 144 x + 27 x^2 -  2x^3}{108 (1- x)^4} - \frac{x (13 + 17 x)}{18 (1-x)^5}\ln x,\\
%
g^{(1)}_4 (x, y ) &= - \frac{ x^2 y  \ln x  }{3  (x-y)^3 (1 - x)^3} \left\{ x^2 ( 5 + 7 x ) + y \left[ 2 + 7 (x - 3) x \right]\right\}\non
&-  \frac{y^2 x \ln y}{ 3  (y-x)^3 (1 - y)^3 } \left\{ y^2 (5 + 7 y) + x \left[2 + 7 (y -3 ) y \right] \right\}\non
& + \frac{x y }{3 (1- x)^2 (1 - y)^2 (x-y)^2 }\left( x + y  - 13 x^2 - 13 y^2  + 8 x y + 15 x^2 y  + 15 x y^2 - 14 x^2 y^2 \right),
\eeq
\beq
%
g^{(1)}_5 (x, y ) &= -  \frac{ x^2 y  \ln x  }{9  (x-y)^3 (1 - x)^3} \left[ x^2 ( 11 +  x ) +  (x -5  )  (x + 2) y \right]\non
&-  \frac{y^2 x \ln y}{ 9  (y-x)^3(1 -y)^3 } \left[ y^2 (11+  y) +  (y-5)(y +2  ) x\right] \non
& - \frac{x y }{9 (1 - x)^2 (1 - y)^2 (x-y)^2 }\left( 5 x + 5 y  + 7 x^2+ 7 y^2   - 32 x y + 3 x^2 y + 3 x y^2 + 2 x^2 y^2 \right).
\eeq

\subsubsection{Wino and Higgsino contributions}
\label{App:winohiggsino}

The loop functions $g^{(1)}_{\tilde{g} \tilde{w}}$, $g^{(1)}_{\tilde{w}} (x)$, $f_1 (x)$ and $f_3(x , y )$ are given by
\beq
g^{(1)}_{\tilde{g} \tilde{w}} (x,y) &= - \sqrt{x y }\biggl[  \frac{x \ln x}{(x-y) (1- x)^4} + \frac{ y \ln y }{(y-x) (1- y)^4} \non
& + \frac{11 - 7 ( x + y) + 2 ( x^2  +y^2)  - 10 x y + 5  x y (x  +  y)  - x^2 y^2 }{ 6 (1-x)^3 (1-y)^3 }\biggr] \non
 & - \frac{x^2 \ln x}{ 2 (x-y)(1 - x)^4 }- \frac{y^2 \ln y}{ 2 (y-x)(1 - y)^4 }\non
 & - \frac{2 + 5 (x+  y)  -( x^2 + y^2) - 22 x y + 5 x y ( x   +  y)  + 2 x^2 y^2}{12 (1 - x)^3  (1 - y)^3},\\
g^{(1)}_{\tilde{w}} (x) &= \frac{ -5 - 67 x - 13 x^2 + x^3}{12 (1-x)^4} - \frac{x (3 + 4 x )}{(1 - x)^5} \ln x,\\
 f_1 (x) & = - \frac{ x + 1}{ 4 (1 - x)^2} - \frac{x}{ 2 (1 - x)^3 } \ln x ,\\
f_3(x , y ) & = - \frac{x^2 [x (1 + x + y ) - 3 y]}{ (x-y)^3(1 - x)^3 } \ln x 
- \frac{y^2 [y (1 + x + y ) - 3 x]}{ (y-x)^3 (1-y)^3} \ln y \non
& ~~ - 2 \frac{x^2 + y^2 - x y - x^2 y - x y^2 + x^2 y^2 }{(1-x)^2 (1-y)^2 (x-y)^2},\\
f_3 (x) & = \frac{{x^2 - 8 x - 17}}{6 (1 - x)^4 } - \frac{ 3 x + 1 }{ (1-x)^5} \ln x,
\label{eq:f3x}
\eeq
where 
$\displaystyle \lim_{y \to x } f_3 (x , y ) = f_3 (x)$.\footnote{
We found that in eq.~(A.15) in ref.~\cite{Altmannshofer:2009ne}, $f_3(x) = (x^2 - 6x - 17 )/[6 (1-x)^4 ] - (3x + 1 ) \ln x / (1 -x)^5$ should be replaced by  eq.~\eqref{eq:f3x}.
}

\bibliographystyle{utphys}
\bibliography{ref}

\providecommand{\href}[2]{#2}\begingroup\raggedright\begin{thebibliography}{10}

\bibitem{Hamzaoui:1998nu}
C.~Hamzaoui, M.~Pospelov, and M.~Toharia, ``{Higgs mediated FCNC in
  supersymmetric models with large $tan \beta$},''
  \href{http://dx.doi.org/10.1103/PhysRevD.59.095005}{{\em Phys. Rev.}
  {\bfseries D59} (1999) 095005},
\href{http://arxiv.org/abs/hep-ph/9807350}{{\ttfamily arXiv:hep-ph/9807350
  [hep-ph]}}.

\bibitem{Babu:1999hn}
K.~S. Babu and C.~F. Kolda, ``{Higgs mediated $B^0 \to \mu^{+} \mu^{-}$ in
  minimal supersymmetry},''
  \href{http://dx.doi.org/10.1103/PhysRevLett.84.228}{{\em Phys. Rev. Lett.}
  {\bfseries 84} (2000) 228--231},
\href{http://arxiv.org/abs/hep-ph/9909476}{{\ttfamily arXiv:hep-ph/9909476
  [hep-ph]}}.

\bibitem{Chankowski:2000ng}
P.~H. Chankowski and L.~Slawianowska, ``{$B^0_{d,s} \to \mu^- \mu^+$ decay in
  the MSSM},'' \href{http://dx.doi.org/10.1103/PhysRevD.63.054012}{{\em Phys.
  Rev.} {\bfseries D63} (2001) 054012},
\href{http://arxiv.org/abs/hep-ph/0008046}{{\ttfamily arXiv:hep-ph/0008046
  [hep-ph]}}.

\bibitem{Bobeth:2001sq}
C.~Bobeth, T.~Ewerth, F.~Kruger, and J.~Urban, ``{Analysis of neutral Higgs
  boson contributions to the decays $\bar{B}$( $s^{)} \to \ell^{+} \ell^{-}$
  and $\bar{B} \to K \ell^{+} \ell^{-}$},''
  \href{http://dx.doi.org/10.1103/PhysRevD.64.074014}{{\em Phys. Rev.}
  {\bfseries D64} (2001) 074014},
\href{http://arxiv.org/abs/hep-ph/0104284}{{\ttfamily arXiv:hep-ph/0104284
  [hep-ph]}}.

\bibitem{Isidori:2001fv}
G.~Isidori and A.~Retico, ``{Scalar flavor changing neutral currents in the
  large $\tan \beta$ limit},''
  \href{http://dx.doi.org/10.1088/1126-6708/2001/11/001}{{\em JHEP} {\bfseries
  11} (2001) 001},
\href{http://arxiv.org/abs/hep-ph/0110121}{{\ttfamily arXiv:hep-ph/0110121
  [hep-ph]}}.

\bibitem{Isidori:2002qe}
G.~Isidori and A.~Retico, ``{$B_{s,d} \to \ell^{+} \ell^{-}$ and $K_{L} \to
  \ell^{+} \ell^{-}$ in SUSY models with nonminimal sources of flavor
  mixing},'' \href{http://dx.doi.org/10.1088/1126-6708/2002/09/063}{{\em JHEP}
  {\bfseries 09} (2002) 063},
\href{http://arxiv.org/abs/hep-ph/0208159}{{\ttfamily arXiv:hep-ph/0208159
  [hep-ph]}}.

\bibitem{Crivellin:2010er}
A.~Crivellin, ``{Effective Higgs Vertices in the generic MSSM},''
  \href{http://dx.doi.org/10.1103/PhysRevD.83.056001}{{\em Phys. Rev.}
  {\bfseries D83} (2011) 056001},
\href{http://arxiv.org/abs/1012.4840}{{\ttfamily arXiv:1012.4840 [hep-ph]}}.

\bibitem{Crivellin:2011jt}
A.~Crivellin, L.~Hofer, and J.~Rosiek, ``{Complete resummation of
  chirally-enhanced loop-effects in the MSSM with non-minimal sources of
  flavor-violation},'' \href{http://dx.doi.org/10.1007/JHEP07(2011)017}{{\em
  JHEP} {\bfseries 07} (2011) 017},
\href{http://arxiv.org/abs/1103.4272}{{\ttfamily arXiv:1103.4272 [hep-ph]}}.

\bibitem{Crivellin:2012zz}
A.~Crivellin and C.~Greub, ``{Two-loop supersymmetric QCD corrections to
  Higgs-quark-quark couplings in the generic MSSM},''
  \href{http://dx.doi.org/10.1103/PhysRevD.87.015013,
  10.1103/PhysRevD.87.079901}{{\em Phys. Rev.} {\bfseries D87} (2013) 015013},
  \href{http://arxiv.org/abs/1210.7453}{{\ttfamily arXiv:1210.7453 [hep-ph]}}.
[Erratum: Phys. Rev.D87,079901(2013)].

\bibitem{Choudhury:1998ze}
S.~R. Choudhury and N.~Gaur, ``{Dileptonic decay of $B_s$ meson in SUSY models
  with large $\tan \beta$},''
  \href{http://dx.doi.org/10.1016/S0370-2693(99)00203-8}{{\em Phys. Lett.}
  {\bfseries B451} (1999) 86--92},
\href{http://arxiv.org/abs/hep-ph/9810307}{{\ttfamily arXiv:hep-ph/9810307
  [hep-ph]}}.

\bibitem{Huang:2000sm}
C.-S. Huang, W.~Liao, Q.-S. Yan, and S.-H. Zhu, ``{$B_s \to$ lepton + lepton -
  in a general 2 HDM and MSSM},''
  \href{http://dx.doi.org/10.1103/PhysRevD.64.059902,
  10.1103/PhysRevD.63.114021}{{\em Phys. Rev.} {\bfseries D63} (2001) 114021},
  \href{http://arxiv.org/abs/hep-ph/0006250}{{\ttfamily arXiv:hep-ph/0006250
  [hep-ph]}}.
[Erratum: Phys. Rev.D64,059902(2001)].

\bibitem{Xiong:2001up}
Z.~Xiong and J.~M. Yang, ``{$B$ meson dileptonic decays enhanced by
  supersymmetry with large $\tan\beta$},''
  \href{http://dx.doi.org/10.1016/S0550-3213(02)00091-3}{{\em Nucl. Phys.}
  {\bfseries B628} (2002) 193--216},
\href{http://arxiv.org/abs/hep-ph/0105260}{{\ttfamily arXiv:hep-ph/0105260
  [hep-ph]}}.

\bibitem{Dedes:2001fv}
A.~Dedes, H.~K. Dreiner, and U.~Nierste, ``{Correlation of $B_s \to \mu^{+}
  \mu^{-}$ and (g-2) ($\mu$) in minimal supergravity},''
  \href{http://dx.doi.org/10.1103/PhysRevLett.87.251804}{{\em Phys. Rev. Lett.}
  {\bfseries 87} (2001) 251804},
\href{http://arxiv.org/abs/hep-ph/0108037}{{\ttfamily arXiv:hep-ph/0108037
  [hep-ph]}}.

\bibitem{Bobeth:2002ch}
C.~Bobeth, T.~Ewerth, F.~Kruger, and J.~Urban, ``{Enhancement of B(anti-B($d$)
  $\to \mu^{+} \mu^{-)}$ / B(anti-B($s$) $\to \mu^{+} \mu^{-)}$ in the MSSM
  with minimal flavor violation and large tan beta},''
  \href{http://dx.doi.org/10.1103/PhysRevD.66.074021}{{\em Phys. Rev.}
  {\bfseries D66} (2002) 074021},
\href{http://arxiv.org/abs/hep-ph/0204225}{{\ttfamily arXiv:hep-ph/0204225
  [hep-ph]}}.

\bibitem{Baek:2002rt}
S.~Baek, P.~Ko, and W.~Y. Song, ``{Implications on SUSY breaking mediation
  mechanisms from observing $B_s \to \mu^+ \mu^-$ and the muon (g-2)},''
  \href{http://dx.doi.org/10.1103/PhysRevLett.89.271801}{{\em Phys. Rev. Lett.}
  {\bfseries 89} (2002) 271801},
\href{http://arxiv.org/abs/hep-ph/0205259}{{\ttfamily arXiv:hep-ph/0205259
  [hep-ph]}}.

\bibitem{Dedes:2002zx}
A.~Dedes, H.~K. Dreiner, U.~Nierste, and P.~Richardson, ``{Trilepton events and
  $B_s \to \mu^{+} \mu^-$ : No lose for mSUGRA at the Tevatron?},''
\href{http://arxiv.org/abs/hep-ph/0207026}{{\ttfamily arXiv:hep-ph/0207026
  [hep-ph]}}.

\bibitem{Mizukoshi:2002gs}
J.~K. Mizukoshi, X.~Tata, and Y.~Wang, ``{Higgs mediated leptonic decays of
  $B_s$ and $B_d$ mesons as probes of supersymmetry},''
  \href{http://dx.doi.org/10.1103/PhysRevD.66.115003}{{\em Phys. Rev.}
  {\bfseries D66} (2002) 115003},
\href{http://arxiv.org/abs/hep-ph/0208078}{{\ttfamily arXiv:hep-ph/0208078
  [hep-ph]}}.

\bibitem{Baek:2002wm}
S.~Baek, P.~Ko, and W.~Y. Song, ``{SUSY breaking mediation mechanisms and (g-2)
  ($\mu$), $B \to X_{s} \gamma$, $B \to X_{s} \ell^{+} \ell^{-}$ and $B_s \to
  \mu^{+} \mu^{-}$},''
  \href{http://dx.doi.org/10.1088/1126-6708/2003/03/054}{{\em JHEP} {\bfseries
  03} (2003) 054},
\href{http://arxiv.org/abs/hep-ph/0208112}{{\ttfamily arXiv:hep-ph/0208112
  [hep-ph]}}.

\bibitem{Ecker:1991ru}
G.~Ecker and A.~Pich, ``{The Longitudinal muon polarization in $K_L\to \mu^+
  \mu^-$},''
\href{http://dx.doi.org/10.1016/0550-3213(91)90056-4}{{\em Nucl. Phys.}
  {\bfseries B366} (1991) 189--205}.

\bibitem{Isidori:2003ts}
G.~Isidori and R.~Unterdorfer, ``{On the short distance constraints from
  $K_{L,S} \to \mu^+ \mu^-$},''
  \href{http://dx.doi.org/10.1088/1126-6708/2004/01/009}{{\em JHEP} {\bfseries
  01} (2004) 009},
\href{http://arxiv.org/abs/hep-ph/0311084}{{\ttfamily arXiv:hep-ph/0311084
  [hep-ph]}}.

\bibitem{DAmbrosio:2017klp}
G.~D'Ambrosio and T.~Kitahara, ``{Direct $CP$ Violation in $K \to \mu^+
  \mu^-$},'' \href{http://dx.doi.org/10.1103/PhysRevLett.119.201802}{{\em Phys.
  Rev. Lett.} {\bfseries 119} no.~20, (2017) 201802},
\href{http://arxiv.org/abs/1707.06999}{{\ttfamily arXiv:1707.06999 [hep-ph]}}.

\bibitem{LHCb:KsMuMu}
R.~Aaij {\em et~al.}, ``Improved limit on the branching fraction of the rare
  decay $k^{0}_{S}\to \mu^{+}\mu^{-}$,''
  \href{http://dx.doi.org/10.1140/epjc/s10052-017-5230-x}{{\em The European
  Physical Journal C} {\bfseries 77} no.~10, (Oct, 2017) 678}.

\bibitem{DMS_FPCP}
D.~Martinez~Santos.
\newblock
  \url{https://cds.cern.ch/record/2270191/files/fpcp2017-MartinezSantos.pdf}.
  LHCb-TALK-2017-164, at FPCP 2017.

\bibitem{MassInsertion}
L.~Hall, V.~Kostelecky, and S.~Raby, ``New flavor violations in supergravity
  models,''
  \href{http://dx.doi.org/https://doi.org/10.1016/0550-3213(86)90397-4}{{\em
  Nuclear Physics B} {\bfseries 267} no.~2, (1986) 415 -- 432}.

\bibitem{Altmannshofer:2009ne}
W.~Altmannshofer, A.~J. Buras, S.~Gori, P.~Paradisi, and D.~M. Straub,
  ``{Anatomy and Phenomenology of FCNC and CPV Effects in SUSY Theories},''
  \href{http://dx.doi.org/10.1016/j.nuclphysb.2009.12.019}{{\em Nucl. Phys.}
  {\bfseries B830} (2010) 17--94},
\href{http://arxiv.org/abs/0909.1333}{{\ttfamily arXiv:0909.1333 [hep-ph]}}.

\bibitem{Rosiek:1995kg}
J.~Rosiek, ``{Complete set of Feynman rules for the MSSM: Erratum},''
\href{http://arxiv.org/abs/hep-ph/9511250}{{\ttfamily arXiv:hep-ph/9511250
  [hep-ph]}}.

\bibitem{Allanach:2008qq}
B.~C. Allanach {\em et~al.}, ``{SUSY Les Houches Accord 2},''
  \href{http://dx.doi.org/10.1016/j.cpc.2008.08.004}{{\em Comput. Phys.
  Commun.} {\bfseries 180} (2009) 8--25},
\href{http://arxiv.org/abs/0801.0045}{{\ttfamily arXiv:0801.0045 [hep-ph]}}.

\bibitem{Crivellin:2017gks}
A.~Crivellin, G.~D'Ambrosio, T.~Kitahara, and U.~Nierste, ``{$K\to \pi
  \nu\overline{\nu}$ in the MSSM in light of the
  $\epsilon^{\prime}_K/\epsilon_K$ anomaly},''
  \href{http://dx.doi.org/10.1103/PhysRevD.96.015023}{{\em Phys. Rev.}
  {\bfseries D96} no.~1, (2017) 015023},
\href{http://arxiv.org/abs/1703.05786}{{\ttfamily arXiv:1703.05786 [hep-ph]}}.

\bibitem{Blum:2011ng}
T.~Blum {\em et~al.}, ``{The $K\to(\pi\pi)_{I=2}$ Decay Amplitude from Lattice
  QCD},'' \href{http://dx.doi.org/10.1103/PhysRevLett.108.141601}{{\em Phys.
  Rev. Lett.} {\bfseries 108} (2012) 141601},
\href{http://arxiv.org/abs/1111.1699}{{\ttfamily arXiv:1111.1699 [hep-lat]}}.

\bibitem{Blum:2012uk}
T.~Blum {\em et~al.}, ``{Lattice determination of the $K \to (\pi\pi)_{I=2}$
  Decay Amplitude $A_2$},''
  \href{http://dx.doi.org/10.1103/PhysRevD.86.074513}{{\em Phys. Rev.}
  {\bfseries D86} (2012) 074513},
\href{http://arxiv.org/abs/1206.5142}{{\ttfamily arXiv:1206.5142 [hep-lat]}}.

\bibitem{Blum:2015ywa}
T.~Blum {\em et~al.}, ``{$K \rightarrow \pi\pi$ $\Delta I=3/2$ decay amplitude
  in the continuum limit},''
  \href{http://dx.doi.org/10.1103/PhysRevD.91.074502}{{\em Phys. Rev.}
  {\bfseries D91} no.~7, (2015) 074502},
\href{http://arxiv.org/abs/1502.00263}{{\ttfamily arXiv:1502.00263 [hep-lat]}}.

\bibitem{Bai:2015nea}
{\bfseries RBC, UKQCD} Collaboration, Z.~Bai {\em et~al.}, ``{Standard Model
  Prediction for Direct CP Violation in $K \to \pi \pi$ Decay},''
  \href{http://dx.doi.org/10.1103/PhysRevLett.115.212001}{{\em Phys. Rev.
  Lett.} {\bfseries 115} no.~21, (2015) 212001},
\href{http://arxiv.org/abs/1505.07863}{{\ttfamily arXiv:1505.07863 [hep-lat]}}.

\bibitem{Pallante:2001he}
E.~Pallante, A.~Pich, and I.~Scimemi, ``{The Standard model prediction for
  $\epsilon^{\prime}_K / \epsilon_K$},''
  \href{http://dx.doi.org/10.1016/S0550-3213(01)00418-7}{{\em Nucl. Phys.}
  {\bfseries B617} (2001) 441--474},
\href{http://arxiv.org/abs/hep-ph/0105011}{{\ttfamily arXiv:hep-ph/0105011
  [hep-ph]}}.

\bibitem{Hambye:2003cy}
T.~Hambye, S.~Peris, and E.~de~Rafael, ``{$\Delta I = 1/2$ and
  $\epsilon^{\prime}_K / \epsilon_K$ in large $N_c$ QCD},''
  \href{http://dx.doi.org/10.1088/1126-6708/2003/05/027}{{\em JHEP} {\bfseries
  05} (2003) 027},
\href{http://arxiv.org/abs/hep-ph/0305104}{{\ttfamily arXiv:hep-ph/0305104
  [hep-ph]}}.

\bibitem{Mullor}
H.~G. Mullor, ``Updated standard model prediction for the kaon direct
  \textit{CP}-violating ratio $\epsilon^{\prime}_k / \epsilon_k$,'' 10, 2017.
\newblock
  \url{https://indico.ific.uv.es/indico/contributionDisplay.py?contribId=146&sessionId=6&confId=2960}.
  Talk given at IX CPAN DAYS.

\bibitem{KLMuMu_theory}
G.~D'Ambrosio, G.~Ecker, G.~Isidori, and H.~Neufeld, ``{Radiative non-leptonic
  kaon decays},'' in {\em {2nd DAPHNE Physics Handbook:265-313}}, pp.~265--313.
\newblock 1994.
\newblock \href{http://arxiv.org/abs/hep-ph/9411439}{{\ttfamily
  arXiv:hep-ph/9411439 [hep-ph]}}.
\newblock
\url{http://preprints.cern.ch/cgi-bin/setlink?base=preprint&categ=cern&id=th-7503-94}.
\newblock

\bibitem{Patrignani:2016xqp}
{\bfseries Particle Data Group} Collaboration, C.~Patrignani {\em et~al.},
  ``{Review of Particle Physics},''
\href{http://dx.doi.org/10.1088/1674-1137/40/10/100001}{{\em Chin. Phys.}
  {\bfseries C40} no.~10, (2016) 100001}.

\bibitem{Jang:2017ieg}
{\bfseries SWME} Collaboration, Y.-C. Jang, W.~Lee, S.~Lee, and J.~Leem,
  ``{Update on $\epsilon_K$ with lattice QCD inputs},'' in {\em {35th
  International Symposium on Lattice Field Theory (Lattice 2017) Granada,
  Spain, June 18-24, 2017}}.
\newblock 2017.
\newblock
\href{http://arxiv.org/abs/1710.06614}{{\ttfamily arXiv:1710.06614 [hep-lat]}}.
\newblock

\bibitem{Endoinprep}
M.~Endo, T.~Goto, T.~Kitahara, S.~Mishima, D.~Ueda, and K.~Yamamoto.
\newblock in preparation.

\bibitem{Kitahara:2016nld}
T.~Kitahara, U.~Nierste, and P.~Tremper, ``{Singularity-free next-to-leading
  order $\Delta$S = 1 renormalization group evolution and
  $\epsilon_K'/\epsilon_K$ in the Standard Model and beyond},''
  \href{http://dx.doi.org/10.1007/JHEP12(2016)078}{{\em JHEP} {\bfseries 12}
  (2016) 078},
\href{http://arxiv.org/abs/1607.06727}{{\ttfamily arXiv:1607.06727 [hep-ph]}}.

\bibitem{C7_constraints}
S.~Descotes-Genon, L.~Hofer, J.~Matias, and J.~Virto, ``Global analysis of
  $b\to s\ell\ell$ anomalies,''
  \href{http://dx.doi.org/10.1007/JHEP06(2016)092}{{\em Journal of High Energy
  Physics} {\bfseries 2016} no.~6, (Jun, 2016) 92}.

\bibitem{Aaboud:2017sjh}
{\bfseries ATLAS} Collaboration, M.~Aaboud {\em et~al.}, ``{Search for
  additional heavy neutral Higgs and gauge bosons in the ditau final state
  produced in 36 fb$^{-1}$ of $pp$ collisions at $\sqrt{s}$ = 13 TeV with the
  ATLAS detector},''
\href{http://arxiv.org/abs/1709.07242}{{\ttfamily arXiv:1709.07242 [hep-ex]}}.

\bibitem{Buras:1999da}
A.~J. Buras, G.~Colangelo, G.~Isidori, A.~Romanino, and L.~Silvestrini,
  ``{Connections between $\epsilon^{\prime}_K / \epsilon_K$ and rare kaon
  decays in supersymmetry},''
  \href{http://dx.doi.org/10.1016/S0550-3213(99)00645-8}{{\em Nucl. Phys.}
  {\bfseries B566} (2000) 3--32},
\href{http://arxiv.org/abs/hep-ph/9908371}{{\ttfamily arXiv:hep-ph/9908371
  [hep-ph]}}.

\bibitem{Barbieri:1999ax}
R.~Barbieri, R.~Contino, and A.~Strumia, ``{$\epsilon^{\prime}_K$ from
  supersymmetry with nonuniversal $A$ terms?},''
  \href{http://dx.doi.org/10.1016/S0550-3213(99)00747-6}{{\em Nucl. Phys.}
  {\bfseries B578} (2000) 153--162},
\href{http://arxiv.org/abs/hep-ph/9908255}{{\ttfamily arXiv:hep-ph/9908255
  [hep-ph]}}.

\bibitem{Mescia:2006jd}
F.~Mescia, C.~Smith, and S.~Trine, ``{$K_L \to \pi^0 e^+ e^-$ and $K_L \to
  \pi^0 \mu^+ \mu^-$: A Binary star on the stage of flavor physics},''
  \href{http://dx.doi.org/10.1088/1126-6708/2006/08/088}{{\em JHEP} {\bfseries
  08} (2006) 088},
\href{http://arxiv.org/abs/hep-ph/0606081}{{\ttfamily arXiv:hep-ph/0606081
  [hep-ph]}}.

\bibitem{Altmannshofer:2011gn}
W.~Altmannshofer, P.~Paradisi, and D.~M. Straub, ``{Model-Independent
  Constraints on New Physics in $b \to s$ Transitions},''
  \href{http://dx.doi.org/10.1007/JHEP04(2012)008}{{\em JHEP} {\bfseries 04}
  (2012) 008},
\href{http://arxiv.org/abs/1111.1257}{{\ttfamily arXiv:1111.1257 [hep-ph]}}.

\bibitem{Buras:2013uqa}
A.~J. Buras, R.~Fleischer, J.~Girrbach, and R.~Knegjens, ``{Probing New Physics
  with the $B_s \to \mu^+ \mu^-$ Time-Dependent Rate},''
  \href{http://dx.doi.org/10.1007/JHEP07(2013)077}{{\em JHEP} {\bfseries 07}
  (2013) 77},
\href{http://arxiv.org/abs/1303.3820}{{\ttfamily arXiv:1303.3820 [hep-ph]}}.

\bibitem{Crivellin:2017upt}
A.~Crivellin, J.~Heeck, and D.~Mueller, ``{Large $h\to b s$ in generic
  two-Higgs-doublet models},''
\href{http://arxiv.org/abs/1710.04663}{{\ttfamily arXiv:1710.04663 [hep-ph]}}.

\bibitem{Isidori:2004rb}
G.~Isidori, C.~Smith, and R.~Unterdorfer, ``{The Rare decay $K_L\to \pi^0 \mu^+
  \mu^-$ within the SM},''
  \href{http://dx.doi.org/10.1140/epjc/s2004-01879-0}{{\em Eur. Phys. J.}
  {\bfseries C36} (2004) 57--66},
\href{http://arxiv.org/abs/hep-ph/0404127}{{\ttfamily arXiv:hep-ph/0404127
  [hep-ph]}}.

\bibitem{GomezDumm:1998gw}
D.~Gomez~Dumm and A.~Pich, ``{Long distance contributions to the K(L) $\to$
  $\mu^+ \mu^-$ decay width},''
  \href{http://dx.doi.org/10.1103/PhysRevLett.80.4633}{{\em Phys. Rev. Lett.}
  {\bfseries 80} (1998) 4633--4636},
\href{http://arxiv.org/abs/hep-ph/9801298}{{\ttfamily arXiv:hep-ph/9801298
  [hep-ph]}}.

\bibitem{Knecht:1999gb}
M.~Knecht, S.~Peris, M.~Perrottet, and E.~de~Rafael, ``{Decay of pseudoscalars
  into lepton pairs and large N(c) QCD},''
  \href{http://dx.doi.org/10.1103/PhysRevLett.83.5230}{{\em Phys. Rev. Lett.}
  {\bfseries 83} (1999) 5230--5233},
\href{http://arxiv.org/abs/hep-ph/9908283}{{\ttfamily arXiv:hep-ph/9908283
  [hep-ph]}}.

\bibitem{Pich:1995qp}
A.~Pich and E.~de~Rafael, ``{Weak $K$ amplitudes in the chiral and $1/N_c$
  expansions},'' \href{http://dx.doi.org/10.1016/0370-2693(96)00171-2}{{\em
  Phys. Lett.} {\bfseries B374} (1996) 186--192},
\href{http://arxiv.org/abs/hep-ph/9511465}{{\ttfamily arXiv:hep-ph/9511465
  [hep-ph]}}.

\bibitem{Gerard:2005yk}
J.-M. Gerard, C.~Smith, and S.~Trine, ``{Radiative kaon decays and the penguin
  contribution to the $\Delta I = 1/2$ rule},''
  \href{http://dx.doi.org/10.1016/j.nuclphysb.2005.09.040}{{\em Nucl. Phys.}
  {\bfseries B730} (2005) 1--36},
\href{http://arxiv.org/abs/hep-ph/0508189}{{\ttfamily arXiv:hep-ph/0508189
  [hep-ph]}}.

\bibitem{Gorbahn:2006bm}
M.~Gorbahn and U.~Haisch, ``{Charm Quark Contribution to $K_L \to \mu^+ \mu^-$
  at Next-to-Next-to-Leading},''
  \href{http://dx.doi.org/10.1103/PhysRevLett.97.122002}{{\em Phys. Rev. Lett.}
  {\bfseries 97} (2006) 122002},
\href{http://arxiv.org/abs/hep-ph/0605203}{{\ttfamily arXiv:hep-ph/0605203
  [hep-ph]}}.

\bibitem{Colangelo:2016ruc}
G.~Colangelo, R.~Stucki, and L.~C. Tunstall, ``{Dispersive treatment of
  $K_S\rightarrow \gamma \gamma $ and $K_S\rightarrow \gamma \ell ^+\ell
  ^-$},'' \href{http://dx.doi.org/10.1140/epjc/s10052-016-4449-2}{{\em Eur.
  Phys. J.} {\bfseries C76} no.~11, (2016) 604},
\href{http://arxiv.org/abs/1609.03574}{{\ttfamily arXiv:1609.03574 [hep-ph]}}.

\bibitem{Colangelo:1998pm}
G.~Colangelo and G.~Isidori, ``{Supersymmetric contributions to rare kaon
  decays: Beyond the single mass insertion approximation},''
  \href{http://dx.doi.org/10.1088/1126-6708/1998/09/009}{{\em JHEP} {\bfseries
  09} (1998) 009},
\href{http://arxiv.org/abs/hep-ph/9808487}{{\ttfamily arXiv:hep-ph/9808487
  [hep-ph]}}.

\bibitem{Endo:2016aws}
M.~Endo, S.~Mishima, D.~Ueda, and K.~Yamamoto, ``{Chargino contributions in
  light of recent $\epsilon^{\prime}_K/\epsilon_K$},''
  \href{http://dx.doi.org/10.1016/j.physletb.2016.10.009}{{\em Phys. Lett.}
  {\bfseries B762} (2016) 493--497},
\href{http://arxiv.org/abs/1608.01444}{{\ttfamily arXiv:1608.01444 [hep-ph]}}.

\bibitem{Buras:2015yba}
A.~J. Buras, M.~Gorbahn, S.~Jäger, and M.~Jamin, ``{Improved anatomy of
  $\epsilon^{\prime}_K/\epsilon_K$ in the Standard Model},''
  \href{http://dx.doi.org/10.1007/JHEP11(2015)202}{{\em JHEP} {\bfseries 11}
  (2015) 202},
\href{http://arxiv.org/abs/1507.06345}{{\ttfamily arXiv:1507.06345 [hep-ph]}}.

\bibitem{Buras:2015xba}
A.~J. Buras and J.-M. Gérard, ``{Upper bounds on
  $\epsilon^{\prime}_K/\epsilon_K$ parameters B$_{6}^{(1/2)}$ and
  B$_{8}^{(3/2)}$ from large N QCD and other news},''
  \href{http://dx.doi.org/10.1007/JHEP12(2015)008}{{\em JHEP} {\bfseries 12}
  (2015) 008},
\href{http://arxiv.org/abs/1507.06326}{{\ttfamily arXiv:1507.06326 [hep-ph]}}.

\bibitem{Buras:2016fys}
A.~J. Buras and J.-M. Gerard, ``{Final state interactions in $K\rightarrow \pi
  \pi $ decays: $\Delta I=1/2$ rule vs. $\epsilon^{\prime}_K/\epsilon_K$},''
  \href{http://dx.doi.org/10.1140/epjc/s10052-016-4586-7}{{\em Eur. Phys. J.}
  {\bfseries C77} no.~1, (2017) 10},
\href{http://arxiv.org/abs/1603.05686}{{\ttfamily arXiv:1603.05686 [hep-ph]}}.

\bibitem{Lellouch:2000pv}
L.~Lellouch and M.~Luscher, ``{Weak transition matrix elements from finite
  volume correlation functions},''
  \href{http://dx.doi.org/10.1007/s002200100410}{{\em Commun. Math. Phys.}
  {\bfseries 219} (2001) 31--44},
\href{http://arxiv.org/abs/hep-lat/0003023}{{\ttfamily arXiv:hep-lat/0003023
  [hep-lat]}}.

\bibitem{Colangelo:2001df}
G.~Colangelo, J.~Gasser, and H.~Leutwyler, ``{$\pi \pi$ scattering},''
  \href{http://dx.doi.org/10.1016/S0550-3213(01)00147-X}{{\em Nucl. Phys.}
  {\bfseries B603} (2001) 125--179},
\href{http://arxiv.org/abs/hep-ph/0103088}{{\ttfamily arXiv:hep-ph/0103088
  [hep-ph]}}.

\bibitem{GarciaMartin:2011cn}
R.~Garcia-Martin, R.~Kaminski, J.~R. Pelaez, J.~Ruiz~de Elvira, and F.~J.
  Yndurain, ``{The Pion-pion scattering amplitude. IV: Improved analysis with
  once subtracted Roy-like equations up to 1100 MeV},''
  \href{http://dx.doi.org/10.1103/PhysRevD.83.074004}{{\em Phys. Rev.}
  {\bfseries D83} (2011) 074004},
\href{http://arxiv.org/abs/1102.2183}{{\ttfamily arXiv:1102.2183 [hep-ph]}}.

\bibitem{Colangelo:NA62}
G.~Colangelo.
\newblock Talk given at the NA62 Physics Handbook MITP Workshop.

\bibitem{Kitahara:2016otd}
T.~Kitahara, U.~Nierste, and P.~Tremper, ``{Supersymmetric Explanation of $CP$
  Violation in $K\to \pi\pi$ Decays},''
  \href{http://dx.doi.org/10.1103/PhysRevLett.117.091802}{{\em Phys. Rev.
  Lett.} {\bfseries 117} no.~9, (2016) 091802},
\href{http://arxiv.org/abs/1604.07400}{{\ttfamily arXiv:1604.07400 [hep-ph]}}.

\bibitem{Kagan:1999iq}
A.~L. Kagan and M.~Neubert, ``{Large $\Delta I = 3/2$ contribution to
  $\epsilon^{\prime}_K / \epsilon$ in supersymmetry},''
  \href{http://dx.doi.org/10.1103/PhysRevLett.83.4929}{{\em Phys. Rev. Lett.}
  {\bfseries 83} (1999) 4929--4932},
\href{http://arxiv.org/abs/hep-ph/9908404}{{\ttfamily arXiv:hep-ph/9908404
  [hep-ph]}}.

\bibitem{Cirigliano:2003nn}
V.~Cirigliano, A.~Pich, G.~Ecker, and H.~Neufeld, ``{Isospin violation in
  $\epsilon^{\prime}_K$},''
  \href{http://dx.doi.org/10.1103/PhysRevLett.91.162001}{{\em Phys. Rev. Lett.}
  {\bfseries 91} (2003) 162001},
\href{http://arxiv.org/abs/hep-ph/0307030}{{\ttfamily arXiv:hep-ph/0307030
  [hep-ph]}}.

\bibitem{Cirigliano:2003gt}
V.~Cirigliano, G.~Ecker, H.~Neufeld, and A.~Pich, ``{Isospin breaking in $K \to
  \pi \pi$ decays},'' \href{http://dx.doi.org/10.1140/epjc/s2003-01579-3}{{\em
  Eur. Phys. J.} {\bfseries C33} (2004) 369--396},
\href{http://arxiv.org/abs/hep-ph/0310351}{{\ttfamily arXiv:hep-ph/0310351
  [hep-ph]}}.

\bibitem{Bailey:2015tba}
{\bfseries SWME} Collaboration, J.~A. Bailey, Y.-C. Jang, W.~Lee, and S.~Park,
  ``{Standard Model evaluation of $\epsilon_K$ using lattice QCD inputs for
  $\hat{B}_K$ and $V_{cb}$},''
  \href{http://dx.doi.org/10.1103/PhysRevD.92.034510}{{\em Phys. Rev.}
  {\bfseries D92} no.~3, (2015) 034510},
\href{http://arxiv.org/abs/1503.05388}{{\ttfamily arXiv:1503.05388 [hep-lat]}}.

\bibitem{Amhis:2016xyh}
Y.~Amhis {\em et~al.}, ``{Averages of $b$-hadron, $c$-hadron, and $\tau$-lepton
  properties as of summer 2016},''
\href{http://arxiv.org/abs/1612.07233}{{\ttfamily arXiv:1612.07233 [hep-ex]}}.

\bibitem{Bigi:2017njr}
D.~Bigi, P.~Gambino, and S.~Schacht, ``{A fresh look at the determination of
  $|V_{cb}|$ from $B\to D^{*} \ell \nu$},''
  \href{http://dx.doi.org/10.1016/j.physletb.2017.04.022}{{\em Phys. Lett.}
  {\bfseries B769} (2017) 441--445},
\href{http://arxiv.org/abs/1703.06124}{{\ttfamily arXiv:1703.06124 [hep-ph]}}.

\bibitem{Grinstein:2017nlq}
B.~Grinstein and A.~Kobach, ``{Model-Independent Extraction of $|V_{cb}|$ from
  $\bar{B}\rightarrow D^* \ell \overline{\nu}$},''
  \href{http://dx.doi.org/10.1016/j.physletb.2017.05.078}{{\em Phys. Lett.}
  {\bfseries B771} (2017) 359--364},
\href{http://arxiv.org/abs/1703.08170}{{\ttfamily arXiv:1703.08170 [hep-ph]}}.

\bibitem{Bernlochner:2017xyx}
F.~U. Bernlochner, Z.~Ligeti, M.~Papucci, and D.~J. Robinson, ``{Tensions and
  correlations in $|V_{cb}|$ determinations},''
  \href{http://dx.doi.org/10.1103/PhysRevD.96.091503}{{\em Phys. Rev.}
  {\bfseries D96} no.~9, (2017) 091503},
\href{http://arxiv.org/abs/1708.07134}{{\ttfamily arXiv:1708.07134 [hep-ph]}}.

\bibitem{Bevan:2013kaa}
A.~Bevan {\em et~al.}, ``{Standard Model updates and new physics analysis with
  the Unitarity Triangle fit},''
\href{http://dx.doi.org/10.1016/j.nuclphysbps.2013.06.015}{{\em Nucl. Phys.
  Proc. Suppl.} {\bfseries 241-242} (2013) 89--94}.

\bibitem{Ambrosino:2006ek}
{\bfseries KLOE} Collaboration, F.~Ambrosino {\em et~al.}, ``{Determination of
  CP and CPT violation parameters in the neutral kaon system using the
  Bell-Steinberger relation and data from the KLOE experiment},''
  \href{http://dx.doi.org/10.1088/1126-6708/2006/12/011}{{\em JHEP} {\bfseries
  12} (2006) 011},
\href{http://arxiv.org/abs/hep-ex/0610034}{{\ttfamily arXiv:hep-ex/0610034
  [hep-ex]}}.

\bibitem{Crivellin:2010ys}
A.~Crivellin and M.~Davidkov, ``{Do squarks have to be degenerate? Constraining
  the mass splitting with Kaon and D mixing},''
  \href{http://dx.doi.org/10.1103/PhysRevD.81.095004}{{\em Phys. Rev.}
  {\bfseries D81} (2010) 095004},
\href{http://arxiv.org/abs/1002.2653}{{\ttfamily arXiv:1002.2653 [hep-ph]}}.

\bibitem{Gabbiani:1996hi}
F.~Gabbiani, E.~Gabrielli, A.~Masiero, and L.~Silvestrini, ``{A Complete
  analysis of FCNC and $CP$ constraints in general SUSY extensions of the
  standard model},'' \href{http://dx.doi.org/10.1016/0550-3213(96)00390-2}{{\em
  Nucl. Phys.} {\bfseries B477} (1996) 321--352},
\href{http://arxiv.org/abs/hep-ph/9604387}{{\ttfamily arXiv:hep-ph/9604387
  [hep-ph]}}.

\bibitem{Buras:2010pza}
A.~J. Buras, D.~Guadagnoli, and G.~Isidori, ``{On $\epsilon_K$ Beyond Lowest
  Order in the Operator Product Expansion},''
  \href{http://dx.doi.org/10.1016/j.physletb.2010.04.017}{{\em Phys. Lett.}
  {\bfseries B688} (2010) 309--313},
\href{http://arxiv.org/abs/1002.3612}{{\ttfamily arXiv:1002.3612 [hep-ph]}}.

\bibitem{Garron:2016mva}
{\bfseries RBC/UKQCD} Collaboration, N.~Garron, R.~J. Hudspith, and A.~T.
  Lytle, ``{Neutral Kaon Mixing Beyond the Standard Model with $n_f=2+1$ Chiral
  Fermions Part 1: Bare Matrix Elements and Physical Results},''
  \href{http://dx.doi.org/10.1007/JHEP11(2016)001}{{\em JHEP} {\bfseries 11}
  (2016) 001},
\href{http://arxiv.org/abs/1609.03334}{{\ttfamily arXiv:1609.03334 [hep-lat]}}.

\bibitem{Bagger:1997gg}
J.~A. Bagger, K.~T. Matchev, and R.-J. Zhang, ``{QCD corrections to flavor
  changing neutral currents in the supersymmetric standard model},''
  \href{http://dx.doi.org/10.1016/S0370-2693(97)00920-9}{{\em Phys. Lett.}
  {\bfseries B412} (1997) 77--85},
\href{http://arxiv.org/abs/hep-ph/9707225}{{\ttfamily arXiv:hep-ph/9707225
  [hep-ph]}}.

\bibitem{Ipanema}
D.~M. Santos, P.~Álvarez Cartelle, M.~Borsato, V.~G. Chobanova, J.~G.
  Pardinñas, M.~L. Martínez, and M.~R. Pernas, ``Ipanema-$\beta$: tools and
  examples for hep analysis on gpu,''
  \href{http://arxiv.org/abs/1706.01420}{{\ttfamily arXiv:1706.01420
  [hep-ex]}}.

\bibitem{IEEE}
J.~Brest, S.~Greiner, B.~Boskovic, M.~Mernik, and V.~Zumer, ``{Self-Adapting
  Control Parameters in Differential Evolution: A Comparative Study on
  Numerical Benchmark Problems},''
\href{http://dx.doi.org/10.1109/TEVC.2006.872133}{{\em IEEE Transactions on
  Evolutionary Computation} {\bfseries 10} (2006) 646--657}.

\bibitem{Planck}
P.~A.~R. Ade {\em et~al.}, ``Planck 2015 results xiii. cosmological
  parameters,'' \href{http://dx.doi.org/110.1051/0004-6361/201525830}{{\em
  Astronomy \& Astrophysics} {\bfseries 594} no.~A13, (Oct, 2016) }.

\bibitem{Costa:2017gup}
J.~C. Costa {\em et~al.}, ``{Likelihood Analysis of the Sub-GUT MSSM in Light
  of LHC 13-TeV Data},''
\href{http://arxiv.org/abs/1711.00458}{{\ttfamily arXiv:1711.00458 [hep-ph]}}.

\bibitem{Bagnaschi:2017tru}
E.~Bagnaschi {\em et~al.}, ``{Likelihood Analysis of the pMSSM11 in Light of
  LHC 13-TeV Data},''
\href{http://arxiv.org/abs/1710.11091}{{\ttfamily arXiv:1710.11091 [hep-ph]}}.

\bibitem{Bagnaschi:2015eha}
E.~A. Bagnaschi {\em et~al.}, ``{Supersymmetric Dark Matter after LHC Run 1},''
  \href{http://dx.doi.org/10.1140/epjc/s10052-015-3718-9}{{\em Eur. Phys. J.}
  {\bfseries C75} (2015) 500},
\href{http://arxiv.org/abs/1508.01173}{{\ttfamily arXiv:1508.01173 [hep-ph]}}.

\bibitem{Bagnaschi:2016xfg}
E.~Bagnaschi {\em et~al.}, ``{Likelihood Analysis of the Minimal AMSB Model},''
  \href{http://dx.doi.org/10.1140/epjc/s10052-017-4810-0}{{\em Eur. Phys. J.}
  {\bfseries C77} no.~4, (2017) 268},
\href{http://arxiv.org/abs/1612.05210}{{\ttfamily arXiv:1612.05210 [hep-ph]}}.

\bibitem{Cuoco:2017iax}
A.~Cuoco, J.~Heisig, M.~Korsmeier, and M.~Krämer, ``{Constraining heavy dark
  matter with cosmic-ray antiprotons},''
\href{http://arxiv.org/abs/1711.05274}{{\ttfamily arXiv:1711.05274 [hep-ph]}}.

\bibitem{Hisano:2006nn}
J.~Hisano, S.~Matsumoto, M.~Nagai, O.~Saito, and M.~Senami, ``{Non-perturbative
  effect on thermal relic abundance of dark matter},''
  \href{http://dx.doi.org/10.1016/j.physletb.2007.01.012}{{\em Phys. Lett.}
  {\bfseries B646} (2007) 34--38},
\href{http://arxiv.org/abs/hep-ph/0610249}{{\ttfamily arXiv:hep-ph/0610249
  [hep-ph]}}.

\bibitem{Ibe:2012sx}
M.~Ibe, S.~Matsumoto, and R.~Sato, ``{Mass Splitting between Charged and
  Neutral Winos at Two-Loop Level},''
  \href{http://dx.doi.org/10.1016/j.physletb.2013.03.015}{{\em Phys. Lett.}
  {\bfseries B721} (2013) 252--260},
\href{http://arxiv.org/abs/1212.5989}{{\ttfamily arXiv:1212.5989 [hep-ph]}}.

\bibitem{Harnik:2002vs}
R.~Harnik, D.~T. Larson, H.~Murayama, and A.~Pierce, ``{Atmospheric neutrinos
  can make beauty strange},''
  \href{http://dx.doi.org/10.1103/PhysRevD.69.094024}{{\em Phys. Rev.}
  {\bfseries D69} (2004) 094024},
\href{http://arxiv.org/abs/hep-ph/0212180}{{\ttfamily arXiv:hep-ph/0212180
  [hep-ph]}}.

\end{thebibliography}\endgroup

\end{document}